\documentclass[a4paper, onecolumn, accepted=2022-08-28]{quantumarticle}

\pdfoutput=1

\usepackage{amsfonts,amsmath,amssymb,amsthm}
\usepackage{mathtools, thmtools,thm-restate}
\usepackage{braket}
\usepackage{graphicx}
\usepackage{caption}
\usepackage{enumitem}
\usepackage{tabularx}
\usepackage{tcolorbox}
\usepackage{pdfpages}

\usepackage[colorlinks=true,allcolors=blue]{hyperref}

\DeclareMathOperator{\Ima}{Im}
\newcommand{\bigO}[1]{\mathcal{O}\left( #1 \right)}
\newcommand{\bOt}[1]{\widetilde{\mathcal O}\left(#1\right)}
\newcommand{\Tr}{\mbox{\rm Tr}}
\newcommand{\tr}[1]{\Tr\left(#1\right)}

\newtheorem{theorem}{Theorem} 
\newtheorem{proposition}[theorem]{Proposition}

\makeatletter
\AtBeginDocument{\let\LS@rot\@undefined}
\makeatother

\begin{document}

\title{Towards Quantum Advantage via Topological Data Analysis}

\author{Casper Gyurik}
\email[]{c.f.s.gyurik@liacs.leidenuniv.nl}
\affiliation{LIACS, Leiden University, Niels Bohrweg 1, 2333 CA Leiden, Netherlands}

\author{Chris Cade}
\affiliation{QuSoft, Centrum Wiskunde \& Informatica (CWI), Science Park 123, 1098 XG Amsterdam, Netherlands}

\author{Vedran Dunjko}
\affiliation{LIACS, Leiden University, Niels Bohrweg 1, 2333 CA Leiden, Netherlands}
\affiliation{LION, Leiden University, Niels Bohrweg 2, 2333 CA Leiden, Netherlands}

\date{August 28th, 2022}

\begin{abstract}
  Even after decades of quantum computing development, examples of generally useful quantum algorithms with exponential speedups over classical counterparts are scarce. 
  Recent progress in quantum algorithms for linear-algebra positioned quantum machine learning (QML) as a potential source of such useful exponential improvements.
  Yet, in an unexpected development, a recent series of ``dequantization'' results has equally rapidly removed the promise of exponential speedups for several QML algorithms.
  This raises the critical question whether exponential speedups of other linear-algebraic QML algorithms persist.
  In this paper, we study the quantum-algorithmic methods behind the algorithm for topological data analysis of Lloyd, Garnerone and Zanardi through this lens.
  We provide evidence that the problem solved by this algorithm is classically intractable by showing that its natural generalization is as hard as simulating the one clean qubit model -- 
  which is widely believed to require superpolynomial time on a classical computer --
  and is thus very likely immune to dequantizations.
  Based on this result, we provide a number of new quantum algorithms for problems such as rank estimation and complex network analysis, along with complexity-theoretic evidence for their classical intractability.
  Furthermore, we analyze the suitability of the proposed quantum algorithms for near-term implementations.
  Our results provide a number of useful applications for full-blown, and restricted quantum computers with a guaranteed exponential speedup over classical methods, recovering some of the potential for linear-algebraic QML to become one of quantum computing's killer applications.
\end{abstract}

\maketitle

\section{Introduction
  \label{sec:intro}}

Quantum machine learning (QML) is a rapidly growing 
field~\cite{dunjko:review, biamonte:review} that has brought forth numerous proposals regarding ways for quantum computers to help analyze data.
Several of these proposals involve using quantum algorithms for linear algebra -- most notably Harrow, Hassidim and Lloyd's matrix inversion algorithm~\cite{harrow:hhl} -- to exponentially speed up tasks in machine learning.
Other proposals such as the use of parameterized quantum circuits~\cite{havlivcek:pqcs, schuld:pqcs, benedetti:pqcs} provide a different approach based on identifying genuinely new quantum learning models (rather than speedups of established methods), which are more amenable to near-term quantum computing restrictions.
These QML proposals have all been hailed as possible examples of quantum computing's ``killer application'': genuinely and broadly useful quantum algorithms which superpolynomially outperform their best known classical counterparts (which are very rare even if full-blown quantum computing is assumed).

However, previously speculated superpolynomial speedups of linear-algebraic QML proposals were revealed to actually be at most polynomial speedups, as exponentially faster classical algorithms were devised that operate under analogous assumptions~\cite{tang:dequantization, chia:dequantizations}.
Nevertheless, practically relevant polynomial speedups may persist~\cite{kerenidis:qrs, lloyd:qpca}. 
While quadratic speedups have obvious appeal on paper, recent analysis involving concrete near-term device properties revealed that low-degree polynomial improvements are not expected to translate to real-world advantages due to various overheads~\cite{babbush:poly_speedup}.
Thus, finding  superpolynomial speedups is of great importance, especially in the early days of practical quantum computing. 
Consequently, it is imperative to re-examine other linear-algebraic QML algorithms to ensure that speculated superpolynomial quantum speedups will not be lost due to development of better classical algorithms.

In this paper, we focus on the quantum-algorithmic methods used by the comparatively less studied algorithm for topological data analysis (TDA) of Lloyd, Garnerone and Zanardi (LGZ)~\cite{lloyd:lgz_algorithm}, and on the TDA problem itself.
We show that the underlying linear-algebraic methods are ``safe'' against general dequantization approaches of the type introduced in~\cite{tang:dequantization, chia:dequantizations}, and that the corresponding computational problem is generally classically intractable (under widely-believed complexity-theoretic assumptions).
This further establishes the potential of these methods to be a source of useful quantum algorithms with superpolynomial speedups over classical methods, which we concretely demonstrate by connecting them to practical problems in machine learning and complex network analysis.
Additionally, we discuss the possibilities of near-term implementations of these quantum methods, which helps position TDA and related problems in the domain of NISQ~\cite{preskill:nisq} devices as well.
The main contributions of this paper are as follows:
\vspace{-3pt}
\begin{itemize}[leftmargin=7pt, itemsep=0.5pt]
\item We provide evidence that TDA (as solved by the LGZ algorithm) is classically intractable. Specifically, we show that a generalization of the TDA problem is as hard as simulating the one clean qubit model of quantum computation, which is widely believed to require superpolynomial time on a classical computer.
\item We provide efficient quantum algorithms for rank estimation and complex network analysis based on the quantum algorithmic methods underlying the LGZ algorithm, along with complexity-theoretic evidence for the classical hardness of the underlying problems.
\item We analyze the possibilities and challenges of near-term implementations of the quantum-algorithmic methods of the LGZ algorithm, focusing on providing several techniques to reduce the required resources, making it more suitable for low-qubit computations. 
\end{itemize} 

We note that while our results do not imply that the narrow TDA problem as solved by the algorithm of LGZ is itself classically intractable (our generalization, however, is shown to be classically intractable), they do eliminate the possibility of a generic dequantization method that does not take into account the specifics of the TDA problem (as is also the case for our extension to complex network analysis).
Nonetheless, our results show that the extension to rank estimation is fully classically intractable, resulting in a provable superpolynomial quantum speedup for this practical problem.
To analyze whether it is possible to further strengthen the argument for quantum advantage (or, to actually find an efficient classical algorithm) for the narrow TDA problem, we closely investigate the state-of-the-art classical algorithms and we highlight the significant theoretical hurdles that, at least currently, stymie such classical approaches.

\medskip

The paper is organized as follows.
For didactic purposes we provide in Section~\ref{sec:tda} a detailed description of the quantum algorithm of LGZ and the background on topological data analysis.
Our main results on the underlying classical hardness are presented in Section~\ref{sec:quantum_advantage}.
In Section~\ref{sec:beyond_betti} we discuss how to extend the applicability of the methods used by the quantum algorithm of LGZ, and we discuss the potential for near-term implementations in Section~\ref{sec:nisq}.
We finish the paper with a discussion of our results in Section~\ref{sec:summary}.

\section{Topological data analysis and the quantum algorithm for Betti number estimation
  \label{sec:tda}}

Topological data analysis is a recent approach to data analysis that extracts robust features from a dataset by inferring properties of the shape of the data.
This is perhaps best explained in analogy to a better-known method: much like how principal component analysis extracts features (i.e., the singular values characterizing the spread of the data in the directions of highest variance) that are invariant under translation and rotation of the data, topological data analysis goes a step further and extract features that are also invariant under bending and stretching of the data (i.e., by inferring properties of its general shape).
Because of this invariance of the extracted features, topological data analysis techniques are inherently robust to noise in the data.

The theory behind topological data analysis is fairly extensive, but most of it we will not need for our purpose.
Namely, we can set most of the topology aside and tackle the issue in linear-algebraic terms, which are well-suited for quantum approaches.
In this section we introduce the relevant linear-algebraic concepts, and we briefly review the quantum algorithm for topological data analysis of Lloyd, Garnerone and Zanardi (LGZ)~\cite{lloyd:lgz_algorithm}.

\subsection{Background and definitions
  \label{subsec:background&defs}}

\begin{figure*}
\includegraphics[width=0.825\textwidth]{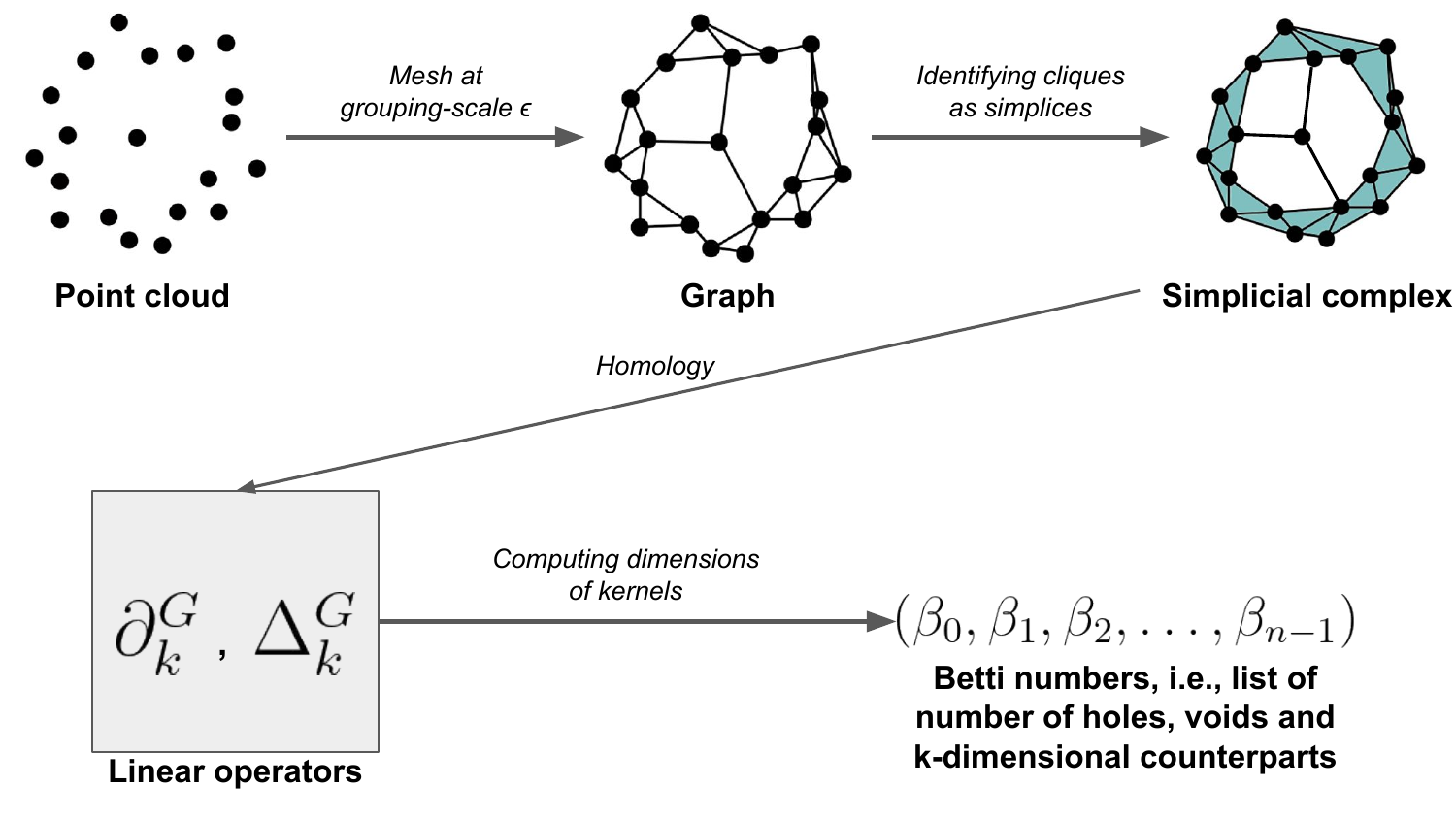}%
\centering
\captionsetup{justification=raggedright}
\caption{\label{fig:tda} The pipeline of topological data analysis (adapted from~\cite{govek:figure}).
First, points that are within $\epsilon$ distance are connected to create a graph.
Afterwards, cliques in this graph are identified with simplices to create a simplicial complex.
Next, homology is used to construct linear operators that encode the topology.
Finally, the dimensions of the kernels of these operators are computed to obtain the Betti numbers which give the number of holes.}
\end{figure*}

In topological data analysis the dataset is typically a point cloud (i.e., a collection of points in some ambient space) and the aim is to extract the shape of the underlying data (i.e., the `source' of these points).
This is done by constructing a connected object -- called a \emph{simplicial complex} -- composed of points, lines, triangles and their higher-dimensional counterparts, whose shape one can study.
After constructing the simplicial complex, features of the shape of the data -- in particular, the number of connected components, holes, voids and higher-dimensional counterparts -- can be extracted using linear-algebraic computations based on \emph{homology}.
An overview of this procedure can be found in Figure~\ref{fig:tda}.

Consider a dataset of points $\{x_i\}_{i=1}^n$ embedded in some space equipped with a distance function $d$ (typically $\mathbb{R}^m$ equipped with the Euclidean distance).
The construction of the simplicial complex from this point cloud proceeds as follows.
First, one constructs a graph by connecting datapoints that are ``close'' to each other.
This is done by choosing a \emph{grouping scale} $\epsilon$ (defining which points are considered ``close'') and connecting all datapoints that are within $\epsilon$ distance from each other.
This yields the graph $G=([n], E_{\epsilon})$, with vertices $[n] \coloneqq \{1, \ldots, n\}$ and edges
\begin{align*}
E_\epsilon = \{(i, j) \mid d(x_i, x_j) \leq \epsilon\}.
\end{align*}
After having constructed this graph, one relates to it a particular kind of simplicial complex called a \emph{clique complex}, by associating its cliques (i.e., complete subgraphs) with the building blocks of a simplicial complex\footnotemark[1].
That is, a 2-clique is considered a line, a 3-clique a triangle, a 4-clique a tetrahedron, and $(k+1)$-cliques the $k$-dimensional counterparts\footnotemark[2].
\footnotetext[1]{The resulting simplicial complex coincides with the Vietoris-Rips complex common in topological data analysis literature~\cite{ghrist:barcodes}.
}
\footnotetext[2]{The shift in the indexing is due to different terminologies in graph theory and topology (e.g., in graph theory a triangle is called a 3-clique, whereas in topology it is called a 2-simplex).}

To fix the notation, let $\text{Cl}_k(G) \subset \{0, 1\}^n$ denote the set of $(k+1)$-cliques in $G$ -- where we encode a subset $\{i_1, \dots, i_k\} \subset [n]$ as an $n$-bit string where the indices $i_k$ specify the positions of the ones in the bitstring -- and let $\chi_k \coloneqq |\text{Cl}_k(G)|$ denote the number of these cliques.
Throughout this paper, we will discuss everything in terms of clique complexes, as this is sufficient for our purposes and allows us to use the more familiar terminology of graph theory.

\begin{figure}
    \centering
    \includegraphics[width=.6\linewidth]{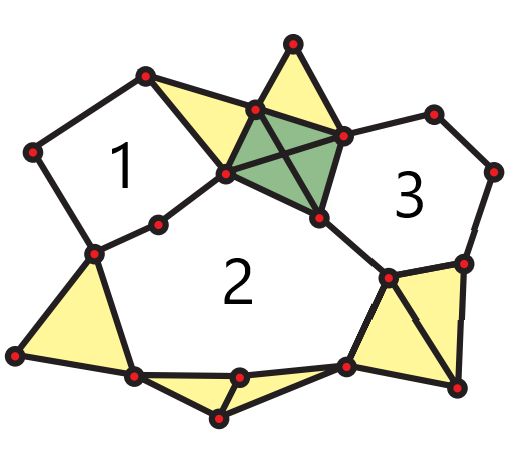}
    \caption{\label{fig:holes} Example of a clique complex with three 1-dimensional holes (adapted from~\cite{ghrist:barcodes}).
    The number of these holes is equal to the first \emph{Betti number}.}
\end{figure}

The constructed clique complex exhibits the features that we want to extract from our dataset -- i.e., the number of $k$-dimensional holes.
For example, in Figure~\ref{fig:holes} we see a clique complex where we can count three 1-dimensional holes.
Interestingly, counting these holes can be done more elegantly using linear algebra by employing constructions from homology.

To extract these features using linear algebra, embed the clique complex into a Hilbert space $\mathcal{H}_k^G$, by raising the set of bitstrings that specify $(k+1)$-cliques to labels of orthonormal basis vectors.
Let $\mathcal{H}_k$ denote the Hilbert space spanned by computational basis states with Hamming weight\footnotemark[3] $k+1$.
Due to the way we encode cliques as bitstrings, we have that $\mathcal{H}_k^G$ is a subspace of  $\mathcal{H}_k$.
\footnotetext[3]{The Hamming weight of a bitstring is the number of 1s in it.}
Moreover, each $\mathcal{H}_k$ is an $\binom{n}{k+1}$-dimensional subspace of the entire $n$-qubit Hilbert space $\mathbb{C}^{2^n}$, and $\mathbb{C}^{2^n} \simeq \bigoplus_{k=-1}^{n-1}\mathcal{H}_k$.

The next step towards extracting features using linear algebra involves studying properties of the \emph{boundary maps} $\partial_k :\mathcal{H}_k  \rightarrow \mathcal{H}_{k-1}$, which are defined by linearly extending the action on the basis states given by
\begin{align}
  \partial_k\ket{j} \coloneqq \sum_{i = 0}^{k}(-1)^i\ket{\widehat{j(i)}},
  \label{eq:boundary_map}
\end{align}
where $\widehat{j(i)}$ denotes the $n$-bit string of Hamming weight~$k$ that encodes the subset obtained by removing the $i$-th element from the subset encoded by $j$ (i.e., we set the $i$-th one in the bitstring $j$ to zero).
By considering the restriction of $\partial_k$ to $\mathcal{H}_k^G$ -- which we denote by~$\partial^G_k$ -- these boundary maps can encode the connectivities of the graph $G$, in which case their image and kernel encode various properties of the corresponding clique complex. 
Intuitively, these boundary maps map a $(k+1)$-clique to a superposition (i.e., a linear combination) of all $k$-cliques that it contains, as seen in Eq.~\eqref{eq:boundary_map}.

These boundary maps allow one to extract features of the shape of a clique complex by studying their images and kernels, and in particular their quotients.
Specifically, the quotient space
\begin{align}
  H_k(G) \coloneqq \ker\partial_k^G /\Ima \partial_{k+1}^G,
  \label{eq:homology_group}
\end{align}
which is called the $k$-th \emph{homology group}, captures features of the shape of the underlying clique complex.
The main feature is the $k$-th \emph{Betti number} $\beta_k^G$, which is defined as the dimension of the $k$-th homology group, i.e., 
\[
\beta_k^G \coloneqq \dim H_k(G).
\]
By construction, the $k$-th Betti number is equal to the number of $k$-dimensional holes in the clique complex.

The main problem in topological data analysis that we study in this paper is the computation of Betti numbers.
To do so, we study the \emph{combinatorial Laplacians}~\cite{eckmann:comb_lapl}, which are defined as
\begin{align}
\label{eq:comb_lapl}
\Delta_k^G= \left(\partial_k^G\right)^\dagger\partial^G_k + \partial_{k+1}^G\left(\partial_{k+1}^G\right)^\dagger.
\end{align}
These combinatorial Laplacians can be viewed as generalized (or rather, higher-order) graph Laplacians in that they encode the connectivity between cliques in the graph as opposed to encoding the connectivity between individual vertices.
We study the combinatorial Laplacians because the discrete version of the Hodge theorem~\cite{eckmann:comb_lapl} tells us that
\begin{align}
  \dim\ker\left(\Delta_k^G\right) = \beta_k^G,
  \label{eq:hodge}
\end{align}
which is often used as a more convenient way to compute Betti numbers~\cite{friedman:computing_betti}, particularly in the case of the quantum algorithm that we discuss in the next section.

In conclusion, if the clique complex is constructed from a point cloud according to the construction discussed above, then computing these Betti numbers can be viewed as a method to extract features of the shape of the data (specifically, the number of holes are present at scale $\epsilon$).
By recording Betti numbers across varying scales $\epsilon$ in a so-called \emph{barcode}~\cite{ghrist:barcodes}, one can discern which holes are ``real'' and which are ``noise'', resulting in feature extraction that is robust to noise in the data.

\subsection{Quantum algorithm for Betti number estimation
\label{subsec:qalg_betti}}
\renewcommand{\thefootnote}{\fnsymbol{footnote}}

The algorithm for Betti number estimation of Lloyd, Garnerone and Zanardi (LGZ)~\cite{lloyd:lgz_algorithm} utilizes Hamiltonian simulation and phase estimation to estimate the dimension of the kernel (i.e., the \emph{nullity}) of the combinatorial Laplacian (which by Eq.~\eqref{eq:hodge} is equal to the corresponding Betti number).
To make our presentation self-contained, we review this quantum algorithm for Betti number estimation (for a more in-depth review see~\cite{gunn:review}).

\begin{figure}[h!]
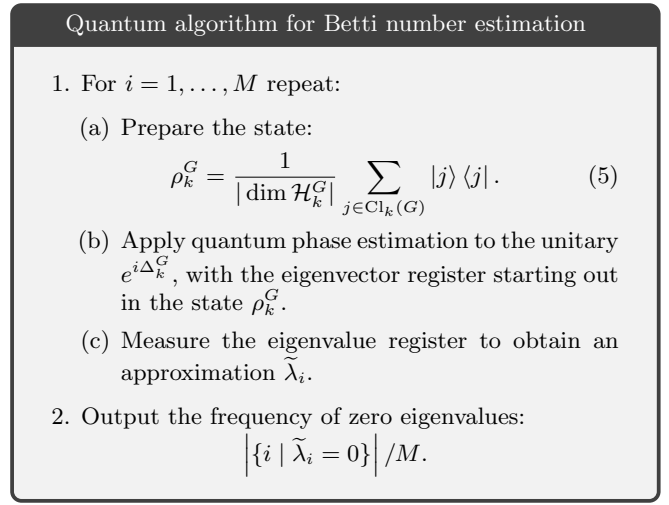

{\hypersetup{citecolor=white}
  \begin{tcolorbox}[title=
    {
       \hspace{-7pt}\mbox{Quantum algorithm for Betti number estimation}
       }]
\begin{enumerate}[leftmargin=*]
\item For $i = 1, \dots, M$ repeat:
  \begin{enumerate}[leftmargin=15pt]
  \item Prepare the state:
  \vspace{-5pt}
    \begin{align}
        \rho_k^G = \frac{1}{|\dim \mathcal{H}_k^G|}\sum_{j \in \text{Cl}_k(G)}\ket{j}\bra{j}.
        \label{eq:mixed_clique}
    \end{align}
  \vspace{-13pt}
  \item Apply quantum phase estimation to the unitary $e^{i\Delta_k^G}$, with the eigenvector register starting out in the state $\rho_k^G$.
  \item Measure the eigenvalue register to obtain an approximation $\widetilde{\lambda}_i$.
  \end{enumerate}
\item Output the frequency of zero eigenvalues:
\vspace{-5pt}
\begin{center}
    $\left|\{i \mid \widetilde{\lambda}_i = 0\}\right| / M$.
\end{center}
\end{enumerate}
\end{tcolorbox}
}

    \caption{Overview of the quantum algorithm of Lloyd, Garnerone and Zanardi (LGZ)~\cite{lloyd:lgz_algorithm}}
    \label{fig:lgz}
\end{figure}

Estimating the nullity of a sparse Hermitian matrix can be achieved using some of the most fundamental quantum-algorithmic primitives. 
Namely, using Hamiltonian simulation and quantum phase estimation one can estimate the eigenvalues of the Hermitian matrix, given that the eigenvector register starts out in an eigenstate.
Moreover, if instead the eigenvector register starts out in the maximally mixed state (which can be thought of as a random choice of an eigenstate), then measurements of the eigenvalue register produce approximations of eigenvalues, sampled uniformly at random from the set of all eigenvalues.
This routine is then repeated to estimate the nullity by simply computing the frequency of zero eigenvalues (recall that the dimension of the kernel is equal to the multiplicity of the zero eigenvalue).
Note that this procedure does not strictly speaking estimate the nullity, but rather the number of small eigenvalues, where the threshold is determined by the precision of the quantum phase estimation (see Section~\ref{subsubsec:approx_betti_numbers} for more details).
The steps of the quantum algorithm for Betti number estimation of LGZ are summarized in Figure~\ref{fig:lgz}.

\medskip 

In Step~1(a), Grover's algorithm is used to prepare the uniform superposition over $\mathcal{H}_k^G$, from which one can prepare the state $\rho_k^G$ by applying a CNOT gate to each qubit of the uniform superposition into some ancilla qubits and tracing those out.
When given access to the adjacency matrix of $G$, one can check in $\bigO{k^2}$ operations whether a bitstring $j \in \{0, 1\}^n$ encodes a valid $k$-clique and mark them accordingly in the application of Grover's algorithm.
By cleverly encoding Hamming weight $k$ strings we can avoid searching over all $n$-bit strings, which requires $\bigO{nk}$ additional gates per round of Grover's algorithm plus an additional one-time cost of $\bigO{n^2k}$~\cite{gunn:review}.
Hence, the runtime of this first step is 
\[
\bigO{n^2k + nk^3\sqrt{\binom{n}{k+1}/\chi_k}} 
\]
where $\chi_k$ denotes the number of $(k+1)$-cliques. 
This runtime is polynomial in the number of vertices~$n$ when
\begin{align}
    \binom{n}{k+1}/\chi_k\in\bigO{\mathrm{poly}(n)}.
    \label{eq:clique_dense}
\end{align}

Throughout this paper we say that a graph is \emph{clique-dense} if it satisfies Eq.~\eqref{eq:clique_dense}.
Note that $\rho^G_k$ can of course also be directly prepared without the use of Grover's algorithm by using rejection sampling:\ choose a subset uniformly at random and accept it if it encodes a valid clique.
This is quadratically less efficient, however it has advantages if one has near-term implementations in mind, as it is a completely classical subroutine.
As we will discuss in more detail in Section~\ref{subsubsec:clique_density}, this state preparation procedure via Grover's algorithm or uniform clique-sampling is a crucial bottleneck in the quantum algorithm.

In Step~1(b), standard methods for Hamiltonian simulation of sparse Hermitian matrices are used together with quantum phase estimation to produce approximations of the eigenvalues of the simulated matrix.
In the original algorithm, the matrix that LGZ simulates in this step is the \emph{Dirac operator}, which is defined as
\begin{align*}
  B_G = \begin{pmatrix}
    0 & \partial^G_1 & 0 & \dots & \dots & 0 \\
    \left(\partial^G_1\right)^\dagger & 0 & \partial^G_2 & \dots & \dots & 0\\
    0 & \left(\partial^G_2\right)^\dagger & 0 & \ddots & \dots & 0 \\
    \vdots & \vdots & \ddots & \ddots & \ddots & \vdots\\
    \vdots & \vdots & \vdots & \ddots&  0 & \partial^G_{n-1} \\
    0 & 0 & 0 & \dots & \left(\partial^G_{n-1}\right)^\dagger & 0 \\
  \end{pmatrix}
\end{align*}
and satisfies
\begin{align}
  B_G^2 =
  \begin{pmatrix}
    \Delta^G_0 & 0        & \dots & 0 \\
    0        & \Delta^G_2 & \dots & 0\\
    \vdots   & \vdots   & \ddots & 0 \\
    0        & 0        & \dots & \Delta^G_{n-1} \\
  \end{pmatrix}.
  \label{eq:dirac_lapl}
\end{align}

From Eq.~\eqref{eq:dirac_lapl} we gather that the probability of obtaining an approximation of an eigenvalue that is equal to zero is proportional to the nullity of the combinatorial Laplacian.
Because $B_G$ is an $n$-sparse Hermitian matrix with entries $0, -1$ and $1$, to which we can implement sparse access using $\bigO{n}$ gates, we can implement $e^{iB}$ using $\bOt{n^2}$ gates~\cite{low:hs} (here $\widetilde{\mathcal{O}}$ suppresses logarithmically growing factors).

We remark that it is also possible to simulate $\Delta_k^G$ directly (as depicted in Figure~\ref{fig:lgz}).
Namely, as $\Delta_k^G$ is an $n^2$-sparse Hermitian matrix whose entries are bounded above by $n$, to which we can implement sparse access using $\bigO{n^4}$ gates (e.g., see Theorem 3.3.4~\cite{goldberg:comb_lapl}), we can implement $e^{i \Delta_k^G}$ using $\bOt{n^6}$ gates~\cite{low:hs}.

The disadvantage to directly simulating $\Delta_k^G$ is that it requires more gates.
However, the advantage is that the Hamiltonian simulation of $\Delta_k^G$ requires fewer qubits compared to the Hamiltonian simulation of $B_G$, namely, $\log\binom{n}{k+1}$ qubits instead of $n$.
Moreover, when the graph is clique-dense one can bypass Step~1(a) by padding $\Delta_k^G$ with all-zero rows and columns and letting the eigenvector start out in the maximally mixed state $I/2^n$ (see Section~\ref{subsubsec:reductions} for more details).

Let $\lambda_{\mathrm{max}}$ denote the largest eigenvalue and let $\lambda_{\mathrm{min}}$ denote the smallest nonzero eigenvalue of $\Delta_k^G$. 
By scaling down the matrix one chooses to simulate (i.e., either $B$ or $\Delta_k^G$) by $1/\lambda_{\mathrm{max}}$ to avoid multiples of $2\pi$, we can tell whether an eigenvalue is equal to zero or not if the precision of the quantum phase estimation is at least $\lambda_{\mathrm{max}}/ \lambda_{\mathrm{min}}$.
By the Gershgorin circle theorem (which states that $\lambda_{\mathrm{max}}$ is bounded above by the maximum sum of absolute values of the entries of a column or row) we know that $\lambda_{\mathrm{max}} \in \bigO{n}$. 
For the general case not much is known in terms of lower bounds on $\lambda_{\mathrm{min}}$.
Nonetheless, even if we do not have such a lower bound, the number of small eigenvalues (as opposed to zero eigenvalues) still conveys topological information about the underlying graph (see Section~\ref{subsubsec:approx_betti_numbers} for more details).
By taking into account the cost of the quantum phase estimation~\cite{n&c}, the total runtime of Step~1(b) becomes~$\bOt{n^3/\lambda_{\mathrm{min}}}$.

Finally, estimating $\beta_k^G/\dim \mathcal{H}_k^G$ up to additive precision $\epsilon$ can be done using $M \in \bigO{\epsilon^{-2}}$ repetitions of Step~1(a) through 1(c).
This brings the total cost of estimating $\beta_k^G/\dim \mathcal{H}_k^G$ up to additive precision~$\epsilon$ to
\begin{align*}
  \bOt{\left(nk^3\sqrt{\binom{n}{k+1}/\chi_k} + n^3/\lambda_{\mathrm{min}} \right)/\epsilon^2}.
\end{align*}

In conclusion, the quantum algorithm for Betti number estimation runs in time polynomial in $n$ under two conditions.
Firstly, the graph has to be clique-dense, i.e., it has to satisfy Eq.~\eqref{eq:clique_dense} (see Section~\ref{subsubsec:clique_density} for more details).
Secondly, the smallest nonzero eigenvalue $\lambda_{\mathrm{min}}$ has to scale at least inverse polynomial in $n$ (see Section~\ref{subsubsec:approx_betti_numbers} for more details).
If both these conditions are satisfied, then the quantum algorithm for Betti number estimation achieves an exponential speedup over the best known classical algorithms if the size of the combinatorial Laplacian -- i.e., the number of $(k+1)$-cliques -- scales exponentially in $n$ (see Section~\ref{subsec:classical_algs} for more details). 

\subsubsection{Approximate Betti numbers
\label{subsubsec:approx_betti_numbers}}

As mentioned in the previous section, the quantum algorithm for Betti number estimation does not strictly speaking estimate the Betti number (i.e., the nullity of the combinatorial Laplacian), but rather the number of small eigenvalues of the combinatorial Laplacian.
This is because little is known in terms of lower bounds for the smallest nonzero eigenvalue of combinatorial Laplacians, and hence it is unclear to what precision one has to estimate the eigenvalues in the quantum phase estimation.
In any case, it is conjectured that for high-dimensional simplicial complexes the smallest nonzero eigenvalue is at least inverse polynomial in $n$~\cite{friedman:computing_betti}, which would imply that quantum phase estimation can in time $\bigO{\mathrm{poly}(n)}$ determine whether an eigenvalue is exactly equal to zero.

Even without knowing a lower bound on the smallest nonzero eigenvalue of the combinatorial Laplacian, we can still perform quantum phase estimation up to some fixed inverse polynomial precision.
The quantum algorithm for Betti number estimation then outputs an estimate of the number of eigenvalues of the combinatorial Laplacian that lie below this precision threshold. 
Throughout this paper we will refer to this as \emph{approximate Betti numbers}, which  turn out to still convey information about the underlying graph.
For instance, Cheeger's inequality -- which relates the sparsest cut of a graph to the smallest nonzero eigenvalues of its standard graph Laplacian -- turns out to have a higher-order generalization that utilizes the combinatorial Laplacian~\cite{gundert:cheeger}.
Moreover, there are several other spectral properties of the combinatorial Laplacian beyond the number of small eigenvalues that also convey topological information about the underlying graph.
Some of these spectral properties can also be efficiently extracted using quantum algorithms (see Section~\ref{subsec:comb_lapl_beyond_betti} for more details).

\subsubsection{Efficient state preparation
  \label{subsubsec:clique_density}}

In Section~\ref{subsec:qalg_betti} we saw that the quantum algorithm for Betti number estimation can efficiently estimate approximate Betti numbers if the input graph satisfies certain criteria.
In particular, the graph has to be such that one can efficiently prepare the maximally mixed state over all its cliques of a given size (i.e., the state in Eq.~\eqref{eq:mixed_clique} in Figure~\ref{fig:lgz}).
In this section we highlight that this state preparation constitutes one of the main bottlenecks in the quantum algorithm for Betti number estimation.

One way to prepare the maximally mixed state over all $k$-cliques of an $n$-vertex graph is to sample $k$-cliques uniformly at random and feed them into the quantum algorithm.
For the quantum algorithm for Betti number estimation to run in time sub-exponential in $n$, we have to be able to sample a $k$-clique uniformly at random in time $n^{o(k)}$.
However, for general graphs finding a $k$-clique cannot be done in time $n^{o(k)}$ unless the exponential time hypothesis fails~\cite{chen:clique}.
Nonetheless, for certain families of graphs, uniform clique sampling can be done much more efficiently, e.g., in time polynomial in $n$ (in which case the quantum algorithm also runs in time polynomial in~$n$).
In particular, the graph's clique-density (i.e., probability that a uniformly random subset of vertices is a clique), or the graph's arboricity (which up to a factor $1/2$ is equivalent to the maximum average degree of a subgraph) are important factors that dictate the efficiency of uniform clique sampling algorithms.
In Section~\ref{subsec:graphs_quantum_advantage} we outline concrete families of graphs (based on their clique-density or arboricity) for which the quantum algorithm achieves a (superpolynomial) speedup over classical algorithms.

\section{Towards quantum advantage for Betti number estimation
\label{sec:quantum_advantage}}
\renewcommand{\thefootnote}{\arabic{footnote}}

In this section we discuss the advantages that the quantum algorithm for Betti number estimation can achieve over classical algorithms.
Firstly, in Section~\ref{subsec:prob_defs}, we precisely delineate and formally define the computational problems that the quantum algorithm for Betti number estimation can (efficiently) solve.
In particular, it is clear that the techniques used in the quantum algorithm for Betti number estimation can also be used to estimate the number of small eigenvalues of arbitrary sparse Hermitian matrix, not just of combinatorial Laplacians.
We take this as the starting point to define our natural generalization, which is called \emph{low-lying spectral density} estimation (a version of which was also studied by Brand\~{a}o~\cite{brandao:thesis}).
Next, in Section~\ref{subsec:hardness_lled}, we show that this generalization is $\mathsf{DQC1}$-hard, which suggests that the quantum-algorithmic methods behind the quantum algorithm for Betti number estimation may be a source of exponential separation between quantum and classical computers.
We also discuss how to potentially close the gap between the topological data analysis problem of Betti number estimation and its generalization, which would show that the topological data analysis problem is itself classically intractable.
Setting aside the complexity theory, in Section~\ref{subsec:classical_algs} we discuss the state-of-the-art classical algorithms for Betti number estimation and compare them with the quantum algorithms for Betti number estimation.
We also discuss promising approaches for developing novel more efficient classical algorithms that take into account the specifics of the combinatorial Laplacian and we clearly delineate the theoretical hurdles that, at least currently, stymie such classical approaches.
After discussing the strengths and weaknesses of the classical algorithms, we identify graphs for which the quantum algorithm can achieve (superpolynomial) speedups over the best known classical algorithms in Section~\ref{subsec:graphs_quantum_advantage}.
\subsection{Problem definitions
\label{subsec:prob_defs}}

In this section we formally define the computational problems whose hardness we will study.
We begin by defining the problems that capture the key steps of the quantum algorithm for Betti number estimation.
Afterwards, we define the problems related to topological data analysis that the quantum algorithm for Betti number estimation aims to solve.
We end this section by discussing the precise relationships between these problems.

The input matrices that we consider are sparse positive semidefinite matrices.
We call a $2^n \times 2^n$ positive semidefinite matrix $\emph{sparse}$ if at most $\bigO{\mathrm{poly}(n)}$ entries in each row are nonzero.
A special class of sparse positive semidefinite matrices that we consider is the class of \emph{log-local} Hamiltonians, i.e., $n$-qubit Hamiltonians that can be written as a sum
\begin{align}
  H = \sum_{j=1}^m H_j,
  \label{eq:local_terms}
\end{align}
where each $H_j$ acts on at most $\bigO{\log n}$ qubits and we assume that $m\in\bigO{\mathrm{poly}(n)}$.

Our problems take as input a specification of a sparse positive semidefinite matrix, and we consider the following two standard cases.
First, we consider the case where the input matrix is specified in terms of \emph{sparse access}.
That is, the input matrix $H \in \mathbb{C}^{2^n \times 2^n}$ is specified by quantum circuits that let us query the values of its entries, and the locations of the nonzero entries.
More precisely, we assume that we are given classical descriptions of $\bigO{\mathrm{poly}(n)}$-sized quantum circuits that implement the oracles $O_H$ and $O_{H, \mathrm{loc}}$, which map
\begin{align*}
  O_H&: \ket{i, j}\ket{0} \mapsto \ket{i, j}\ket{H_{i,j}},\\
  O_{H, \mathrm{loc}}&: \ket{j, \ell}\ket{0} \mapsto \ket{j, \ell}\ket{\nu(j, \ell)},
\end{align*}
where $0 \leq i, j , \ell \leq 2^n - 1$, and $\nu(j, \ell) \in \{0, \dots, 2^n - 1\}$ denotes the location of the $\ell$-th nonzero entry of the $j$-th column of~$H$.
Secondly, for log-local Hamiltonians, we also consider specifying the input matrix $H$ by its \emph{local-terms} $\{H_j\}$ as in Eq.~\eqref{eq:local_terms}.

In order to define the problem of generating approximations of eigenvalues that are sampled uniformly at random, we fix a suitable notion of an approximation of a probability distribution.
In particular, this notion needs to take into account that the algorithm may err on both the estimation of the eigenvalue, and on the probability with which it provides such an estimation.
For this we use the following definition presented in~\cite{wocjan:bqp_complete}.
Let $p$ be some probability distribution over the eigenvalues of a positive semidefinite matrix $H \in \mathbb{C}^{2^n \times 2^n}$.
That is, sampling according to $p$ will output an eigenvalue $\lambda_k$ with probability $p(\lambda_k)$, and $\sum_{k=0}^{2^n -1}p(\lambda_k) = 1$. 
In this context, a probability distribution $q$ with finite support $Y_q \subset \mathbb{R}$ is said to be an \emph{$(\delta, \mu)$-approximation} of $p$ if it satisfies
\[
  \sum_{y \in Y_q\text{ }:\text{ }|y - \lambda_k| < \delta}q(y) \geq (1-\mu)p(\lambda_k),\enskip \forall k \in \{0, \dots, 2^{n-1}\}.
\]
Intuitively, this means that if we draw a sample according to $q$, then this sample will be $\delta$-close to an eigenvalue $\lambda_k$ with probability at least $(1-\mu)p(\lambda_k)$\footnotemark[1].\footnotetext[1]{This definition captures the distribution generated by quantum phase estimation:\ the eigenvector is chosen according to the distribution $p$ specified by the input state, and the output is $\delta$-close to the corresponding eigenvalue with probability at least $(1-\mu)$.}
Using this definition, we define the problem of generating approximations of eigenvalues that are sampled uniformly at random from the set of all eigenvalues as follows.

\vspace{5pt}
\noindent\begin{tabularx}{\linewidth}{l X c}
  \multicolumn{2}{l}{\textbf{\textsf{Sparse uniform eigenvalue sampling (SUES)}}\footnotemark[2]} \\
  \multicolumn{2}{l}{\textbf{Input:}} \\
  1) & A sparse positive semidefinite matrix $H\in\mathbb{C}^{2^n \times 2^n}$, with $||H|| \leq \mathrm{poly}(n)$.\\
  2) & An estimation precision $\delta \in \Omega\left(1/\mathrm{poly}(n)\right)$.\\
  3) & An error probability $\mu \in \Omega\left( 1/\mathrm{poly}(n)\right)$.
\end{tabularx}
\noindent\begin{tabularx}{\linewidth}{l X c}
  \textbf{Output:} & A sample drawn according to a $(\delta, \mu)$-approximation of the uniform distribution over the eigenvalues of $H$.
\end{tabularx}
\footnotetext[2]{In view of noisy quantum computers, it is interesting to consider distributions that are close to these $(\delta,\mu)$-approximation in total variation distance.
Sampling such distributions can be less demanding, however, the precise hardness remains to be analyzed.}

In the quantum algorithm for Betti number estimation, samples from \textsc{sues} are used to estimate the number of eigenvalues of the combinatorial Laplacian that are close to zero.
Clearly, this same idea can be used to estimate the number of eigenvalues that lie in some given interval for arbitrary sparse positive semidefinite matrices.
This is called the \emph{eigenvalue count}~\cite{brandao:thesis}, which for a positive semidefinite matrix $H~\in~\mathbb{C}^{2^n \times 2^n}$ and eigenvalue thresholds $a, b \in \mathbb{R}_{\geq 0}$ is given by
\begin{align*}
  N_H(a, b) = \frac{1}{2^n}\sum_{k\text{ }:\text{ }a \leq \lambda_k\leq b}1,
\end{align*}
where $\lambda_0 \leq \dots \leq \lambda_{2^n-1}$ denote the eigenvalues of $H$.
For a threshold $b \in \Omega\left(1/\mathrm{poly}(n)\right)$, we shall refer to the quantity $N_H(0, b)$ as \emph{low-lying spectral density}.
This precisely captures our notion of the number of eigenvalues close to zero as discussed before.
We define the problem of estimating the low-lying spectral density as follows.

\vspace{3pt}
\noindent\begin{tabularx}{\linewidth}{l X c}
  \multicolumn{2}{l}{\textbf{\textsf{Low-lying spectral density estimation (LLSD)}}\footnotemark[3]} \\
  \multicolumn{2}{l}{\textbf{Input:}}\\
  1) & A sparse positive semidefinite matrix $H \in \mathbb{C}^{2^n \times 2^n}$, with $||H|| \leq \mathrm{poly}(n)$.\\
  2) & A threshold $b \in \Omega\left(1/\mathrm{poly}(n)\right)$.\\
  3) & Precision parameters $\delta, \epsilon \in \Omega\left(1/\mathrm{poly}(n)\right)$.\\
  4) & A success probability $\mu > 1/2$.\\
\end{tabularx}
\noindent\begin{tabularx}{\linewidth}{l X c}
  \textbf{Output:} & An estimate $\chi \in [0,1]$ that, with probability at least $\mu$, satisfies
\end{tabularx}
\vspace{-3pt}
\begin{align*}
    N_H\left(0, b\right) - \epsilon \leq \chi \leq N_H(0, b + \delta) + \epsilon.
\end{align*}

\footnotetext[3]{The exact version of this problem is closely related to $\mathsf{\# P}$~\cite{brown:dos}.}

To provide some intuition behind this definition, note that it is supposed to precisely capture the problem that is solved by repeatedly sampling from \textsc{sues} and computing the frequency of the eigenvalues that lie below the given threshold.
We therefore require the precision parameter~$\delta$ due to the imprecisions in the quantum phase estimation algorithm.
Moreover, the precision parameter~$\epsilon$ is necessary due to the sampling error we incur by estimating a probability by a relative frequency.

Now that we have formally defined the problems that capture the key steps of the quantum algorithm for Betti number estimation, we define the problems related to topological data analysis that they allow us to solve.
For these problems we consider the adjacency matrix of the graph to be the input, as this is usually the input to the quantum algorithm for Betti number estimation.
We define the problem of estimating Betti numbers as follows.

\vspace{5pt}
\noindent\begin{tabularx}{\linewidth}{l X c}
  \multicolumn{2}{l}{\textbf{\textsf{Betti number estimation (BNE)}}\footnotemark[4]} \\
  \multicolumn{2}{l}{\textbf{Input:}}\\
  1) & The adjacency matrix of a graph $G = ([n], E)$.\\
  2) & An integer $0 \leq k \leq n-1.$\\
  3) & A precision parameter $\epsilon \in \Omega\left(1/\mathrm{poly}(n)\right)$.\\
  4) & A success probability $\mu > 1/2$.\\
\end{tabularx}
\noindent\begin{tabularx}{\linewidth}{l X c}
  \textbf{Output:} & An estimate $\chi \in [0,1]$ that, with probability at least $\mu$, satisfies
\end{tabularx}
\vspace{-3pt}
\begin{align*}
   \left|\chi - \frac{\beta_k^G}{\dim \mathcal{H}_k^G}\right| \leq \epsilon.
\end{align*}

\footnotetext[4]{The exact version of this problem is $\mathsf{NP}$-hard~\cite{adamaszek:betti_np}}

As discussed in Section~\ref{subsubsec:approx_betti_numbers}, the quantum algorithm for Betti number estimation does not precisely solve the above problem.
Namely, due to the lack of knowledge regarding lower bounds on the smallest nonzero eigenvalue of the combinatorial Laplacian, we are not always able to estimate the number of eigenvalues that are exactly equal to zero.
Nonetheless, the quantum algorithm for Betti number estimation is still able to estimate the number of eigenvalues of the combinatorial Laplacian that are close to zero, which we called approximate Betti numbers.
We define the problem of estimating approximate Betti numbers as follows.

\vspace{5pt}
\noindent\begin{tabularx}{\linewidth}{l X c}
  \multicolumn{2}{l}{\textbf{\textsf{Approximate Betti number estimation (ABNE)}}} \\
  \multicolumn{2}{l}{\textbf{Input:}}\\
  1) & The adjacency matrix of a graph $G = ([n], E)$.\\
  2) & An integer $0 \leq k \leq n-1.$\\
  3) & Precision parameters $\delta, \epsilon \in \Omega\left(1/\mathrm{poly}(n)\right)$.\\
  4) & A success probability $\mu > 1/2$.\\
\end{tabularx}
\noindent\begin{tabularx}{\linewidth}{l X c}
  \textbf{Output:} & An estimate $\chi \in [0,1]$ that, with probability at least $\mu$, satisfies
\end{tabularx}
\vspace{-3pt}
\begin{align*}
    \frac{\beta_k^G}{\dim \mathcal{H}_k^G} - \epsilon \leq \chi \leq N_{\Delta_k^G}(0, \delta) + \epsilon.
\end{align*}

We are now set to outline the problem that the quantum algorithm for Betti number estimation can efficiently solve.
As discussed in Section~\ref{subsubsec:clique_density}, the quantum algorithm for Betti number estimation can efficiently solve \textsc{abne}, but only in certain regimes.
In particular, one has to be able to efficiently prepare the maximally mixed state over all cliques of a given size from the adjacency matrix of the graph.
As mentioned in Section~\ref{subsubsec:clique_density}, the efficiency of this state preparation depends on the graph's clique-density (i.e., probability that a uniformly random subset of vertices is a clique), or the graph's arboricity (which up to a factor $1/2$ is equivalent to the maximum average degree of a subgraph).
In short, the problem that the quantum algorithm for Betti number estimation can efficiently solve is a restriction of \textsc{abne}, where one is promised that the input graph is such that one can efficiently prepare the maximally mixed state over all cliques of a given size from the adjacency matrix (e.g., if the graph is sufficiently clique-dense or if it has a sufficiently bounded arboricity).
We discuss this in more detail in Section~\ref{subsec:graphs_quantum_advantage}, where we outline sufficient conditions on the graph's clique-density or arboricity that allow the quantum algorithm to efficiently solve \textsc{abne}. 

Next, we will study the complexity of \textsc{llsd} as it is a generalization of the problem that the quantum algorithm for Betti number estimation efficiently solves.
Namely, as we will show in the following section, we can use \textsc{llsd} to directly solve the problem that the quantum algorithm for Betti number estimation efficiently solves.
Note that the input to the quantum algorithm for Betti number estimation is the adjacency matrix, and not the combinatorial Laplacian.
Therefore, in order to use \textsc{llsd} to solve the problem that the quantum algorithm for Betti number estimation efficiently solves, one first has to construct the appropriate input to \textsc{llsd}.
As it is computationally too expensive to enumerate all cliques in your graph, we cannot take the straightforward approach of first computing the combinatorial Laplacian to construct the desired input to \textsc{llsd}. 
Fortunately, we can still use \textsc{llsd} to efficiently solve the problem that the quantum algorithm for Betti number estimation efficiently solves by simulating sparse access to a matrix that is obtained by padding the combinatorial Laplacian with all-zeros columns and rows (see Section~\ref{subsubsec:reductions} for more details).

\subsubsection{Relationships between the problems
\label{subsubsec:reductions}}

\begin{figure*}
\includegraphics[width=0.9\textwidth]{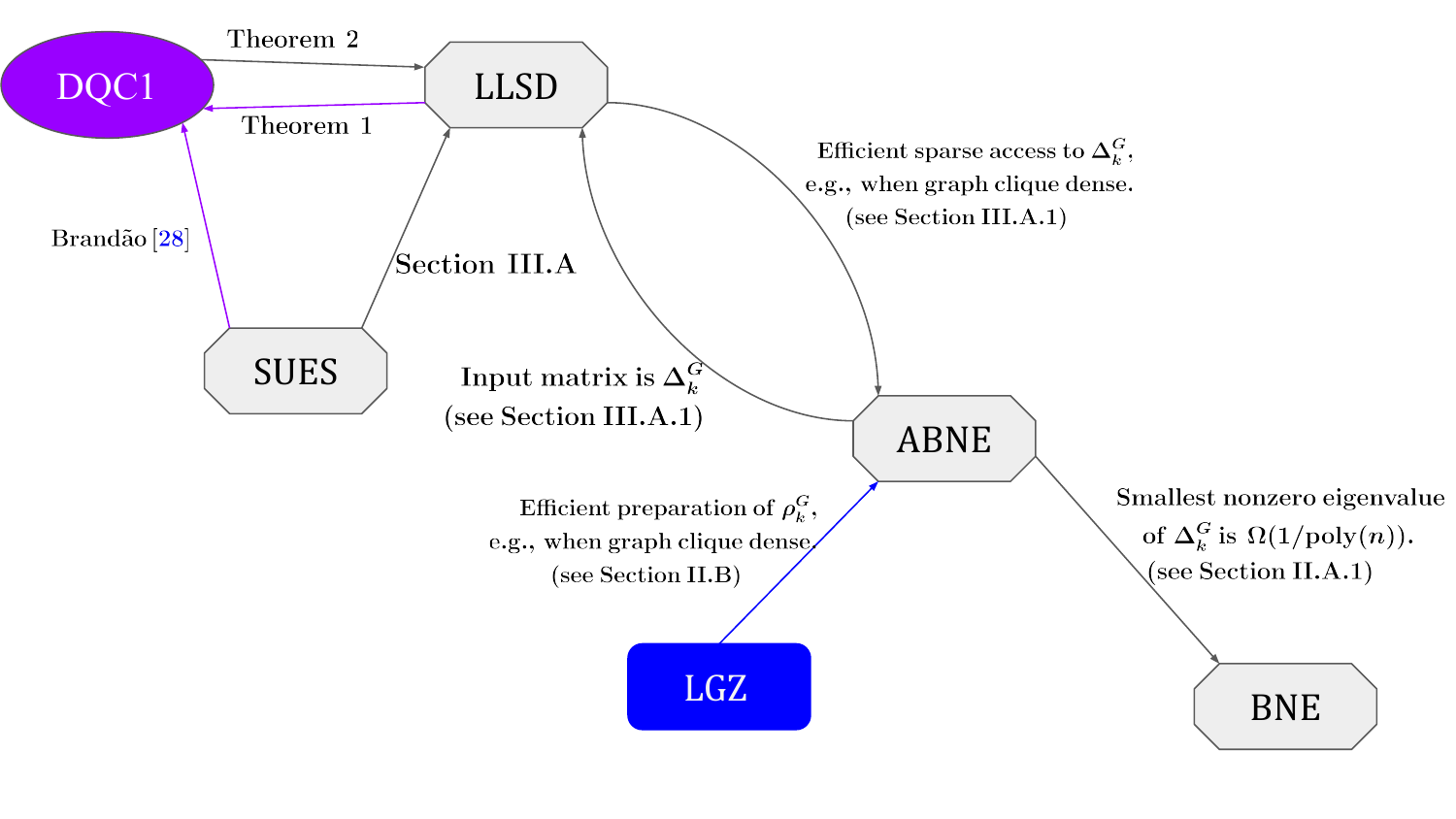}%
\captionsetup{justification=raggedright}
\caption{\label{fig:reductions} Overview of the relations between the problems (octagons), algorithm (rectangle) and complexity class (ellipse) studied in this paper. 
A $\stackrel{\rm C}{\longrightarrow}$ B stand for:\ ``A can efficiently solve B if condition C holds''.
The algorithm studied is that by Lloyd, Garnerone and Zanardi (LGZ) as described in Figure~\ref{fig:lgz}.
The problems are sparse uniform eigenvalue sampling (\textsc{sues}), low-lying spectral density estimation (\textsc{llsd}), approximate Betti number estimation (\textsc{abne}), and Betti number estimation (\textsc{bne}) as defined in Section~\ref{subsec:prob_defs}
The class $\mathsf{DQC1}$ is defined in Section~\ref{subsubsec:dqc1}.}
\end{figure*}

In the previous section we have formally defined the computational problems whose complexity we will study.
In this section we examine the reductions between \textsc{llsd} and the problems related to topological data analysis in order to elucidate the precise relationships.
An overview of the reductions can be found in Figure~\ref{fig:reductions}.

First, we discuss the relationship between \textsc{llsd} and \textsc{abne}.
It is clear that \textsc{llsd} with a combinatorial Laplacian as input produces a solution to the corresponding instance of \textsc{abne}.
It is also clear that \textsc{llsd} can be used to solve \textsc{abne} if given the input of \textsc{abne} (i.e, the adjacency matrix), we can efficiently implement sparse access to a matrix such that an estimate of its low-lying spectral density allows us to recover an estimate of the low-lying spectral density of the combinatorial Laplacian.
Interestingly, it turns out that we can do so if the input graph is clique-dense (i.e., in precisely the regime that is efficiently solvable by the quantum algorithm for Betti number estimation).
Namely, we can efficiently implement sparse access to the $\binom{n}{k+1} \times \binom{n}{k+1}$-sized matrix $\Gamma_k^G$ whose columns and rows are indexed by $(k+1)$-subsets of vertices, and whose entries are given by
\begin{align}
  \left(\Gamma_k^G\right)_{i, j} =
  \begin{cases} (\Delta_k^G)_{i, j} & \text{if $i$ and $j$ are $(k+1)$-cliques,}\\ 0 & \text{otherwise}.
  \end{cases}
\label{eq:gamma_matrix}
\end{align}
In other words, the entries of the columns and rows that correspond to $(k+1)$-cliques are equal to the corresponding entries of the combinatorial Laplacian, and all other entries are equal to zero.
After subtracting the extra nullity caused by adding the $\binom{n}{k+1} - \chi_k$ all-zeros columns and rows, and renormalizing the eigenvalue count by a factor $\binom{n}{k+1}/\chi_k$, the low-lying spectral density of this $\Gamma_k^G$ is equal to the low-lying spectral density of the combinatorial Laplacian.
In equation form, we have that
\begin{align}
  N_{\Delta_k^G}(0, b) = \frac{\binom{n}{k+1}}{\chi_k}N_{\Gamma_k^G}(0, b) - \frac{\binom{n}{k+1} - \chi_k}{\chi_k}.
  \label{eq:llsd_delta_gamma}
\end{align}
From Eq.~\eqref{eq:llsd_delta_gamma}, it is clear that an estimate of $N_{\Gamma_k^G}(0, b)$ up to additive inverse polynomial precision allows us to obtain an estimate of $N_{\Delta_k^G}(0, b)$ up to additive inverse polynomial precision, assuming indeed that the graph is clique-dense -- i.e., that $\chi_k/\binom{n}{k+1} \in \Omega\left(1/\mathrm{poly}(n)\right)$.
Note that this also requires us to have an estimate of $\chi_k/\binom{n}{k+1}$.
Since the graph is clique-dense, it suffices to estimate $\chi_k/\binom{n}{k+1}$ up to additive inverse polynomial precision.
An estimate of $\chi_k/\binom{n}{k+1}$ up to additive error $\epsilon$ can be obtained by drawing $\mathcal{O}(\epsilon^{-2})$ many $k$-subsets of vertices uniformly at random, and computing the fraction of these subsets that constitute an actual $k$-clique.

We emphasize that the above reduction works in precisely the regime where the quantum algorithm for Betti number estimation can efficiently solve \textsc{abne}.
In other words, \textsc{llsd} can be used to directly solve the problem that the quantum algorithm for Betti number estimation can efficiently solve.
In this regard, \textsc{llsd} is indeed a generalization of the problem that the quantum algorithm for Betti number estimation can efficiently solve.

Finally, let us discuss the reductions between \textsc{abne} and \textsc{bne}.
It is clear that \textsc{bne} is reducible to \textsc{abne} if the size of the smallest nonzero eigenvalue of the combinatorial Laplacian is at least inverse polynomial in $n$.
The reverse direction is unclear, as for \textsc{bne} the threshold on the eigenvalues is fixed to be exactly zero.
A possible approach would be to first project the eigenvalues that lie below the given threshold to zero and then count the zero eigenvalues.
However, using techniques inspired by ideas from~\cite{gilyen:block, kitaev:book}, we have only been able to project these eigenvalues close to zero, as opposed to exactly equal to zero, and we are not aware of any way to circumvent this.

\subsection{Classical intractability of low-lying spectral density estimation}
\label{subsec:hardness_lled}

To show that quantum computers have an advantage over classical computers in topological data analysis, one would have to show that Betti number estimation requires exponential time on a classical computer.
In this section we study the classical hardness of the problem efficiently solved by the quantum algorithm for Betti number estimation.
In particular, we show that the natural generalization of this problem (which we called low-lying spectral density estimation) is classically intractable under widely-believed complexity-theoretic assumptions by showing that it is hard for the one clean qubit model of computation.
Afterwards, we discuss how to potentially close the gap between the classical intractability of low-lying spectral density and (approximate) Betti number estimation in order to show that the topological data analysis problem is itself classically intractable. 

\subsubsection{The one clean qubit model of computation}
\label{subsubsec:dqc1}

In the next section we will show that the complexity of the problems defined in Section~\ref{subsec:prob_defs} are closely related to the one clean qubit model of quantum computation~\cite{knill:dqc1}.
In this model we are given a quantum register that is initialized in a state consisting of a single `clean' qubit in the state $\ket{0}$, and $n-1$ qubits in the maximally mixed state.
We can then apply any polynomially-sized quantum circuit to this register, and measure only the first qubit in the computational basis.
Following~\cite{knill:dqc1}, we will refer to the complexity class of problems that can be solved in polynomial time using this model of computation as $\mathsf{DQC1}$ -- ``deterministic quantum computation with a single clean qubit''.

We will refer to a problem as $\mathsf{DQC1}$-hard if any problem in $\mathsf{DQC1}$ can be reduced to it under polynomial time truth-table reductions.
That is, a problem $L$ is $\mathsf{DQC1}$-hard if we can solve any problem in $\mathsf{DQC1}$ using polynomially many nonadaptive queries to an oracle for $L$, together with polynomial time preprocessing of the inputs and postprocessing of the outcomes.
Technically, instead of containing the problem of estimating a given quantity up to additive inverse polynomial precision, $\mathsf{DQC1}$ contains the decision problem of deciding whether this quantity is greater than $1/2 + \sigma$ or less than $1/2 - \sigma$, where $\sigma$ is some inverse polynomial gap.
However, as the estimation versions of these problems are straightforwardly reduced to their decision version using binary search, we will bypass this point from now on and only consider the problems of estimating a given quantity up to inverse polynomial precision~\cite{shor:dqc1}.

It is widely believed that the one clean qubit model of computation is more powerful than classical computation.
For instance, estimating quantities that are supposedly hard to estimate classically, such as the normalized trace of a unitary matrix corresponding to a polynomial-depth quantum circuit and the evaluation of a Jones polynomial at a root of unity, turn out to be complete problems for $\mathsf{DQC1}$~\cite{shor:dqc1}.
Moreover, it has been shown that classical computers cannot efficiently sample from the output distribution of the one clean qubit model up to constant total variation distance error, provided that some complexity theoretic conjectures hold~\cite{tomoyuki:dqc1, tomoyuki:dqc1_k}.

\subsubsection{Hardness of low-lying spectral density estimation for the one clean qubit model
\label{subsec:results}}

Recall that in order to show that quantum computers have an advantage over classical computers in topological data analysis, one would have to show that the problem that the quantum algorithm for Betti number estimation can efficiently solve is hard for classical computers.
In Section~\ref{subsec:prob_defs}, we pointed out that the problem that the quantum algorithm for Betti number estimation can efficiently solve is a restriction of \textsc{abne} to clique-dense graphs (i.e., graphs which satisfy Eq.~\eqref{eq:clique_dense}).
Moreover, we showed in Section~\ref{subsubsec:reductions} that \textsc{llsd} is a generalization of this version of \textsc{abne}.
This motivates us to study the classical hardness of \textsc{llsd}.
In this section we present our results, which show that the complexity of \textsc{llsd} is intimately related to the one clean qubit model.

Our first and main result is that \textsc{llsd} is hard for the class $\mathsf{DQC1}$, even when the input is restricted to log-local Hamiltonians.
As the one clean qubit model of computation is widely believed to be more powerful than classical computation, this shows that \textsc{llsd} is likely hard for classical computers.
We discuss the implications of this result on the classical hardness of the problem that the quantum algorithm for Betti number estimation can efficiently solve in Section~\ref{subsubsec:discussion}.

\begin{theorem}
  \textsc{llsd} is $\mathsf{DQC1}$-hard.
  Moreover, \textsc{llsd} with the input restricted to log-local Hamiltonians remains $\mathsf{DQC1}$-hard.
 \label{thm:dqc1_hardness_llsd}
\end{theorem}

We now give a sketch of our proof of the above theorem, the complete proof can be found in the Supplemental Material.
The main idea behind the proof is to show that we can use \textsc{llsd} to estimate a quantity similar to a normalized subtrace -- or more precisely, a normalized sum of eigenvalues below a given threshold -- which has been shown to be $\mathsf{DQC1}$-hard by Brand\~{a}o~\cite{brandao:thesis}.
We estimate this normalized subtrace by constructing a histogram approximation of the low-lying eigenvalues, and afterwards computing the mean of this histogram.
To construct this histogram, we use \textsc{llsd} to estimate the number of eigenvalues that lie in each bin.
To avoid double counting of eigenvalues due to imprecisions around the thresholds of the bins, we subtract the output of \textsc{llsd} with the eigenvalue threshold set to the lower-threshold of the bin from the output of \textsc{llsd} with the eigenvalue threshold set to the upper-threshold of the bin. 
By doing so, we obtain an estimate of the number of eigenvalues within the bin, and misplace eigenvalues by at most one bin\footnotemark[1].
\footnotetext[1]{
  We believe that our approach could be modified to a Karp reduction by encoding an instance of the normalized subtrace estimation problem into a single instance of \textsc{llsd}.
  This would entail manipulating the Kitaev circuit-to-Hamiltonian construction and taking direct sums of matrices.
  As this reduction is not vital for our claim, we leave this question open for future work.}

Our second result shows that the complexity of \textsc{llsd} is more closely related to $\mathsf{DQC1}$ than just hardness.
Namely, we point out that if the input to \textsc{llsd} is restricted to log-local Hamiltonians (or more generally, any type of Hamiltonian that allows for efficient Hamiltonian simulation using $\bigO{\log(n)}$ ancilla qubits), then it can be solved using the one-clean qubit model.
From this it follows that \textsc{llsd} is $\mathsf{DQC1}$-complete if the input is restricted to log-local Hamiltonians.
The main idea behind why we can solve these instances of \textsc{llsd} using the one clean qubit model is that the one-clean qubit model can simulate having access to up to $\bigO{\log(n)}$ pure qubits~\cite{shor:dqc1}.
These pure qubits allow for Hamiltonian simulation techniques based on the Trotter-Suzuki formula~\cite{lloyd:hs} and for quantum phase estimation up to the required precision.
We summarize this in the following theorem, the proof of which can be found in the Supplemental Material.

\begin{theorem}
  \textsc{llsd} with the input restricted to log-local Hamiltonians is $\mathsf{DQC1}$-complete.
  \label{theorem:complete}
\end{theorem}

As an added result, we find that the complexity of \textsc{sues} with the input restricted to log-local Hamiltonians is also closely related to $\mathsf{DQC1}$.
The complexity of this instance $\textsc{sues}$ was stated as an open problem by Wocjan and Zhang~\cite{wocjan:bqp_complete}.
Moreover, we believe that it is interesting to study the complexity of \textsc{sues}, as this problem can potentially find practical applications beyond both \textsc{llsd} and Betti number estimation.
We remark that \textsc{sues} with the input restricted to log-local Hamiltonians was already shown to be $\mathsf{DQC1}$-hard by Brand\~ao~\cite{brandao:thesis}.
Here we point out that the complexity of this instance of \textsc{sues} is more closely related to the one clean qubit model than just hardness, as it can also be solved using $\mathsf{DQC1}_{\log n}$ circuits, that is, $\mathsf{DQC1}$ circuits where we are allowed to measure logarithmically many of the qubits in the computational basis at the end (to read out the encoding of the eigenvalue).
The proof of the following proposition can be found in the Supplemental Material.

\begin{proposition}
  \textsc{sues} with the input restricted to log-local Hamiltonians can be solved in polynomial time by the one clean qubit model with logarithmically many qubits measured at the end.
  \label{prop:sues}
\end{proposition}

\subsubsection{Closing the gap for classical intractability of $\textsc{abne}$
\label{subsubsec:discussion}}

The results discussed in the previous section are not sufficient to conclude that \textsc{abne} and \textsc{bne} are hard for classical computers, because for these problems the family of input matrices is restricted to combinatorial Laplacians.
Nonetheless, because \textsc{llsd} is a generalization of the problem that the quantum algorithm for Betti number estimation can efficiently solve, our result shows that -- aside from the matter regarding the restriction to combinatorial Laplacians -- the quantum algorithm for Betti number estimation solves a classically intractable problem which in some cases captures interesting information concerning an underlying graph.
Moreover, our result eliminates the possibility of certain routes for dequantization, namely those that are oblivious to the particular structure of the input matrix, which in particular eliminates the approaches of Tang et al.~\cite{chia:dequantizations}.

The open question regarding the classical hardness of \textsc{abne} and the problem that the quantum algorithm for Betti number estimation can efficiently solve is whether \textsc{llsd} remains classically hard when restricted to combinatorial Laplacians of arbitrary or clique-dense graphs, respectively.
Even though these restrictions on the input seem quite stringent, note that our result shows that \textsc{llsd} is already $\mathsf{DQC1}$-hard for the restricted family of log-local Hamiltonians obtained from Kitaev's circuit-to-Hamiltonian construction\footnotemark[1].
Moreover, there exists a family of combinatorial Laplacians that can encode $\mathsf{DQC1}$-hard Hamiltonians, however they are not combinatorial Laplacians of clique complexes~\cite{cade:complexity}.
One way we tried to close this gap  was by investigating whether we could encode Hamiltonians obtained from Kitaev's circuit-to-Hamiltonian construction into combinatorial Laplacians of sufficiently large graphs.
While indeed various matrices related to quantum gates can be found as submatrices of combinatorial Laplacians, we did not succeed in finding an explicit embedding.
In our view, this remains a promising way of showing that \textsc{llsd} remains classically hard when restricted to combinatorial Laplacians (if indeed this claim is true at all).
\footnotetext[1]{In~\cite{cade:schatten_p, brandao:thesis} and our case it is unclear whether this holds for $k$-local Hamiltonians with constant $k$, as the standard constructions of these local Hamiltonians involve a clock register that is too large.}%
\vspace{0pt}

Besides the above approach based on the Kitaev circuit-to-Hamiltonian construction, there are many other constructions that could potentially be used to show that \textsc{llsd} remains classically hard when restricted to combinatorial Laplacians (again, if indeed this claim is true at all).
In particular, there are several constructions used to prove $\mathsf{QMA}$-hardness of the ground-state energy problem for certain families of Hamiltonians (i.e., deciding if the smallest eigenvalue lies above or below some thresholds)\footnotemark[2].
\footnotetext[2]{For an overview of circuit-to-Hamiltonian constructions see~\cite{bookatz:qma}.}%
All of these constructions typically take as input a (verification) circuit and produce a Hamiltonian that has a small eigenvalue if and only if there exists a quantum state (also called a witness) that makes the circuit accept (i.e., if on this input it is more likely to output 1 on the first qubit).
A special property of the Kitaev construction is that for every input to the circuit, there exists a state whose energy with respect to the corresponding Hamiltonian is close to the acceptance probability of the circuit (i.e., not just that there exists small eigenvalue if and only if there exists a state that makes the circuit accept).
This property allowed Brand\~{a}o to prove that normalized sub-trace estimation for these Hamiltonians is $\mathsf{DQC1}$-hard~\cite{brandao:thesis}, which is at the core of our proof of $\mathsf{DQC1}$-hardness of \textsc{llsd}.
Hence, a promising approach to show $\mathsf{DQC1}$-hardness of $\textsc{llsd}$ for a family of Hamiltonians is to look at existing circuit-to-Hamiltonian constructions used to prove $\mathsf{QMA}$-hardness of versions of the ground-state energy problem and investigate whether they also have this special property that the Kitaev construction has (or to see if they can be equipped with it).
This is particularly interesting for the constructions used to show $\mathsf{QMA}$-hardness of the Bose-Hubbard model~\cite{childs:bose}, or the Fermi-Hubbard model~\cite{gorman:fermi}.
The reason for this is that both of these Hamiltonians exhibit similarities to the Hamiltonian of the hardcore fermion model, which is equal to the combinatorial Laplacian of a clique complex.
Specifically, the Hamiltonian $H_G$ of the fermion hardcore model on a graph $G=([n], E)$ is given by
\begin{align}
\label{eq:ham_fermion}
H_G = \sum_{(i, j) \in E}P_i a_i a_j^\dagger P_j + \sum_{i \in V}P_i,
\end{align}
where $P_i = \prod_{(i, j) \in E}(I - n_j)$,  $a_i$ denotes the fermionic annihilation operator, and $n_j$ denotes the fermionic number operator.
For this Hamiltonian $H_G$ it holds that 
\begin{align}
    H_{G} = \bigoplus_{k=0}^{n-1}\Delta_k^{\bar{G}},
    \label{eq:ham_fermion_comb_lapl}
\end{align}
where $\bar{G}$ denotes the complement graph of $G$, and $\Delta_k^G$ denotes the $k$-th combinatorial Laplacian.

Finally, instead of trying to show that the family of combinatorial Laplacians is sufficiently rich, we could also generalize this family of matrices while still remaining relevant to topological data analysis.
For example, one could consider generalizations of combinatorial Laplacians, such as weighted combinatorial Laplacians~\cite{horak:spectra_comb_laplacian} or persistent combinatorial Laplacians~\cite{wang:persistent}, and show that these generalized families are sufficiently rich as to contain $\mathsf{DQC1}$-hard instances.
Besides all the approaches discussed above, other routes such as proving classical hardness of \textsc{llsd} when restricted to other sets of matrices such as $\{0, \pm 1 \}$-matrices, or by going through the discrete structures related to Tutte and Jones polynomials~\cite{ahmadi:tutte,shor:dqc1} could all be possible as well.

The open questions regarding the classical hardness of \textsc{bne} are the same as those regarding the classical hardness of \textsc{abne}, except that there is one additional open question.
Namely, assuming that \textsc{abne} is classically hard, the remaining open question regarding the classical hardness of \textsc{bne} is whether estimating the number of eigenvalues exactly equal to zero is at least as hard as estimating the number of eigenvalues below a given inverse polynomially small threshold.
This question was already addressed in Section~\ref{subsubsec:reductions} when we examined the reductions between \textsc{abne} and \textsc{bne}.
As discussed there, one approach would be to project the eigenvalues below the given threshold to zero, and afterwards count only the zero eigenvalues.

Regardless, even if \textsc{llsd} does not remain classically hard when restricted to combinatorial Laplacians, we can envision practical generalizations of the quantum-algorithmic methods used by the algorithm for Betti number estimation that go beyond Betti numbers, as we will discuss in more detail in Section~\ref{sec:beyond_betti}.
Specifically, in Section~\ref{sec:beyond_betti} we provide efficient quantum algorithms for two concrete examples of such practical generalizations, together with complexity-theoretic evidence of their classical hardness. 
The first example we discuss is numerical rank estimation, an important problem in machine learning, data analysis and many other applications.
The second example is spectral entropy estimation, which can be used as a tool in complex network analysis.

\subsection{Classical algorithms for approximate Betti number estimation
\label{subsec:classical_algs}}

In the previous section we gave complexity-theoretic evidence for quantum advantage in topological data analysis by proving that $\textsc{llsd}$ -- a generalization of \textsc{abne} -- is $\mathsf{DQC1}$-hard.
In this section we will closely investigate the state-of-the-art classical algorithms, to analyze whether it is possible to strengthen the argument for quantum advantage (or, to actually find an efficient classical algorithm) for the topological data analysis problem.
In particular, we will cover classical algorithms based on numerical linear algebra or random walks and analyze the theoretical hurdles that, at least currently, stymie them from performing equally as well as the quantum algorithm. 

To the best of our knowledge, the best known classical algorithms for approximate Betti number estimation is based on a numerical linear algebra algorithm for low-lying spectral density estimation~\cite{ubaru:approximate_rank, cheung:rank, napoli:eigenvalue_counts, lin:spectral_density}.
These algorithms typically run in time linear in the number of nonzero entries.
Since combinatorial Laplacians are $n$-sparse, the number of nonzero entries of the combinatorial Laplacian -- and hence also the runtime of the best known classical algorithm for approximate Betti number estimation -- scales as 
\begin{align*}
    \bigO{n \cdot \chi_k} \in \bigO{n^{k+1}}.
\end{align*}

Recall that the quantum algorithm for Betti number estimation can estimate approximate Betti numbers in time polynomial in $n$ if we can efficiently prepare the maximally mixed state over the cliques of a given size (e.g., if it satisfies Eq.~\eqref{eq:clique_dense}).
For graphs that satisfy this condition, we conclude that the quantum algorithm for Betti number estimation achieves an exponential speedup over the best known classical algorithms if the size of the combinatorial Laplacian -- i.e., the number of $(k+1)$-cliques -- scales exponential in $n$ (which requires $k$ to scale with $n$).
For exponential speedups for Betti number estimation, we also require that the smallest nonzero eigenvalue of the combinatorial Laplacian scales at least inverse polynomially in $n$.

To investigate the actual hardness of approximate Betti number estimation, we go one step further and discuss new possibilities for efficient classical algorithms. 
In particular, we investigate potential classical algorithms that take into account the specifics of the combinatorial Laplacian by using carefully designed random walks.
Firstly, there exists a classical random walk based algorithm that can approximate the spectrum of the $0$th combinatorial Laplacian (i.e., the ordinary graph Laplacian) up to $\epsilon$ distance in the Wasserstein-1 metric in time $\bigO{\mathrm{exp}(1/\epsilon)}$ (i.e., independent of the size of the graph)~\cite{cohen:walk}.
To generalize this to higher-order combinatorial Laplacians, one would have to construct an efficiently implementable walk operator whose spectral properties coincide with the higher-order combinatorial Laplacian.
While potential candidates for such higher-order walk operators have previously been studied~\cite{mukherjee:walk, parzanchevski:walk}, we conclude after substantial literature review that to the best of our knowledge little is known about such higher-order walk operators.
Furthermore, there is no indication that any of the required structure persists from already existing random walk operators.
Note that such a construction must take into account the specifics of the combinatorial Laplacian, since if the construction would work for arbitrary sparse Hermitian matrices, then this would lead to an efficient classical algorithm for \textsc{llsd} (which by Theorem~\ref{thm:dqc1_hardness_llsd} is widely-believed to be impossible).
Finally, even if the methods of~\cite{cohen:walk} are generalized to higher-order combinatorial Laplacians, then the error-scaling of the eigenvalue precision would still be exponentially worse compared to the standard quantum algorithm that combines Hamiltonian simulation and quantum phase estimation.

\subsection{Graphs with quantum speedup
\label{subsec:graphs_quantum_advantage}}

In Section~\ref{subsubsec:clique_density}, we outlined criteria that the graph has to satisfy in order for the quantum algorithm to be able to efficiently estimate (approximate) Betti numbers.
Specifically, the graph has to be such that one can efficiently prepare the input state in Eq.~\eqref{eq:mixed_clique}, e.g., by sampling uniformly at random from cliques of a given size.
Afterwards, in Section~\ref{subsec:classical_algs}, we discussed the best known classical algorithms and we outlined the regimes in which they require superpolynomial runtimes. 
In this section we put these two considerations together and we concretely characterize families of graphs for which the quantum algorithm achieves either a high-degree polynomial, or even a superpolynomial speedup over the best known classical algorithm.
In particular, we identify families of graphs for which the quantum algorithm is efficient and for which the best known classical algorithms are unable to achieve competitive runtimes.

As discussed in Section~\ref{subsubsec:clique_density}, one way to efficiently prepare the input state is to use Grover's algorithm or rejection sampling to sample uniformly at random from cliques of a given size.
Recall that for this to be efficient the graph has to be clique dense, i.e., it has to satisfy Eq.~\eqref{eq:clique_dense}.
To identify a family a clique-dense graphs, let us consider clique sizes $k \geq 3$, let $\gamma > \frac{k-2}{2(k-1)}$ be a constant, and consider any graph on $n$ vertices with at least $\gamma n^2$ edges.
Suppose we want to estimate the $k$-th approximate Betti number of this graph, where $k$ and the precision parameters are constant.
The quantum algorithm for Betti number estimation can do so in time
\begin{align*}
  \bOt{\sqrt{n^{k+1}/\chi_k} + n^3},
\end{align*}
where $\chi_k$ denotes the number of $(k+1)$-cliques. 
Having chosen the graph the way we did, the clique density theorem~\cite{reiher:clique} now directly guarantees that our graph satisfies 
\[
\chi_k \in \Omega(n^{k+1}),
\]
which is a phenomenon known as ``supersaturation''.
In particular, this implies that our graph is clique-dense and that the quantum algorithm for Betti number estimation estimates the required approximate Betti number in time
\[
\bOt{n^3}.
\]
Moreover, as discussed in Section~\ref{subsec:classical_algs}, the best known classical algorithm requires time
\[
    \bigO{n^{k+1}},
\]
as the number of nonzero entries of the corresponding combinatorial Laplacian is at least $\chi_k$.
We conclude that in these instances the quantum algorithm for Betti number estimation achieves a $(k-2)$-degree polynomial speedup over the best known classical methods, which for large enough $k$ might allow for runtime advantages on prospective fault-tolerant computers, even when all overheads are accounted for~\cite{babbush:poly_speedup}.

We can push the separation between the best known classical algorithm and the quantum algorithm even further.
Consider the same setting as above, but with $\gamma = \frac{k-1}{k}$ and we allow $k$ to scale with $n$.
Using a result of Moon and Moser~\cite{moon:clique, lovasz:clique, ugander:clique}, we can derive that in this setting the graph satisfies
\[
\binom{n}{k+1}  / \chi_k \in \bigO{k^{k}}. 
\]
Therefore, the quantum algorithm can estimate the $k$-th approximate Betti number in time 
\[
\bigO{k^{2 + k/2} + n^3}.
\]
On the other hand, the best known classical algorithm runs in time
\[
\bigO{n^{k+1}/k^{2k}},
\]
as the number of nonzero entries of the corresponding combinatorial Laplacian is at least $\chi_k \geq n^{k+1}/k^{2k}$.
In particular, if we let $k$ scale with $n$ in an appropriate way, then the quantum algorithm achieves a superpolynomial speedup over the best known classical method.
For example, if we let the clique size scale as $k \sim \log n$, then the quantum algorithm runs in time 
\[
2^{\bigO{\log n\log \log n}},
\]
whereas the best known classical algorithm runs in time
\[
2^{\bigO{(\log n)^2}},
\]
giving rise to a superpolynomial quantum speedup. 
Note that the graphs in the previous two settings are rather edge-dense (which occurs in topological data analysis if the grouping-scale $\epsilon$ approaches the maximum distance between two datapoints), and it is unknown whether better classical algorithms are possible in this regime.

As also discussed in Section~\ref{subsubsec:clique_density}, besides clique-density another important graph parameter that dictates the runtimes of specialized algorithms for uniform clique sampling is the so-called \emph{arboricity}.
The arboricity of a graph is equivalent (up to a factor $1/2$) to the maximum average degree of a subgraph.
For a graph with $n$ vertices and arboricity $\alpha$, near-optimal classical algorithms sample a $k$-clique uniformly at random in time~\cite{eden:clique}  
\begin{align}
     \bOt{k^k \cdot \max \bigg\{\left(\frac{(n\alpha)^{k/2}}{\chi_k}\right)^{\frac{1}{k-1}}, \enskip \min\Big\{n\alpha, \frac{n\alpha^{k-1}}{\chi_k} \Big\} \bigg\} }.
     \label{eq:runtime_uniform_sampling}
\end{align}

By also considering the algorithm of~\cite{eden:clique}  (i.e., instead of rejection sampling or Grover's algorithm) we strictly expand the family of graphs for which the quantum algorithm achieves a superpolynomial speedup for $\textsc{abne}$.  
In particular, there exists a family of graphs for which the algorithm of~\cite{eden:clique} is superpolynomially more efficient\footnote{We say that a runtime $t_1(n)$ is \emph{superpolynomially more efficient} than a runtime $t_2(n)$ if $\log t_2(n)/\log t_1(n) \rightarrow \infty$ when $n \rightarrow \infty$.} than Grover's algorithm and rejection sampling for the problem of uniform clique sampling.
An example of such a family is as follows:
consider the $n$-vertex graphs consisting of $n/r$ cliques of size $r$ (for simplicity we assume that $n$ is a multiple of $r$), where each $r$-clique is fully-connected with $d$ other $r$-cliques (i.e., all edges between the $2r$ vertices are present).
In other words, consider a $d$-regular graph on $n/r$ vertices, and replace each vertex with an $r$-clique and fully-connect all $r$-cliques that were connected according to the $d$-regular graph we started with. 
Now if we set $d, r = \log n$ and $k = \log\log n$, then the number of $k$-cliques (and thus also the runtime of the best known classical algorithm for \textsc{abne}) scales like $\log(n)^{\log\log(n)}$.
Moreover, the clique-density (and thus also the runtime of rejection sampling and Grover's algorithm) scales like $n^{\log\log(n)}$.
Finally, the runtime of the algorithm of~\cite{eden:clique} scales like $\log\log(n)^{\log\log(n)}$.
In conclusion, for these graphs the algorithm of~\cite{eden:clique} is superpolynomially more efficient than rejection sampling and Grover's algorithm for the problem of uniform clique sampling.
Moreover, for these graphs the quantum algorithm for $\textsc{abne}$ achieves a superpolynomial speedup over the best-known classical algorithm for $\textsc{abne}$, but only if one uses the algorithm of~\cite{eden:clique} (i.e., this speedup goes away if one uses rejection sampling or Grover's algorithm).
We again remark that we are dealing with special types of graphs, and it is unknown whether better classical algorithms are possible in this regime.

\section{Quantum speedups beyond Betti numbers
  \label{sec:beyond_betti}}

In the previous section we provided evidence that the computational problems tackled by the quantum algorithm for Betti number estimation are likely hard for classical computers.
Even though we fell short of showing that the topological data analysis problem of estimating (approximate) Betti number is classically intractable, we did provide evidence that the quantum algorithmic methods that underlie the quantum algorithm for Betti number estimation could give rise to a potential source of practical quantum advantage.
In this section we demonstrate this by discussing extensions of the quantum-algorithmic methods behind the algorithm for Betti number estimation that go beyond Betti numbers.
In particular, we provide efficient quantum algorithms for numerical rank estimation (an important problem in machine learning and data analysis) and spectral entropy estimation (which can be used to compare complex networks), together with complexity-theoretic evidence of their classical hardness.

\subsection{Numerical rank estimation
  \label{subsec:numerical_rank_estimation}}

In this section we identify a practically important application of the problem of estimating the number of small eigenvalues (which we called \textsc{llsd}).
Specifically, we consider the problem of \emph{numerical rank estimation}. 
The numerical rank of a matrix $H \in \mathbb{C}^{2^n \times 2^n}$ is the number of eigenvalues that lie above some given threshold $b$, i.e., it is defined as
\begin{align*}
    r_H(b) = \frac{1}{2^n}\sum_{k\text{ }:\text{ }\lambda_k > b} 1,    
\end{align*}
where $\lambda_1 \leq \dots \leq \lambda_{2^{n}-1}$ denote the eigenvalues of $H$.
By the rank-nullity theorem we have that 
\begin{align*}
r_H(b) = 1 - N_H(0, b),
\end{align*}
which shows that we can estimate the numerical rank using low-lying spectral density estimation and that the error scaling is the same.

Many machine learning and data analysis applications deal with high-dimensional matrices whose relevant information lies in a low-dimensional subspace.
To be specific, it is a standard assumption that the input matrix is the result of adding small perturbations (e.g., noise in the data) to a low-rank matrix.
This small perturbation turns the input matrix into a high-rank matrix, that can be well approximated by a low-rank matrix. 
Techniques such as principle component analysis~\cite{jolliffe:pca} and randomized low-rank approximations~\cite{halko:random_low_rank} are able exploit this property of the input matrix.
However, these techniques often require as input the dimension of this low-dimensional subspace, which is often unknown.
This is where numerical rank estimation comes in, as it can estimate the dimension of the relevant subspace by estimating the number of eigenvalues that lie above the ``noise-threshold''.
In addition, being able to determine whether the numerical rank of a matrix is large or small enables one to assert whether the above low-rank approximation techniques is applicable at all, or not.

From Theorem~\ref{thm:dqc1_hardness_llsd} it directly follows that quantum computers achieve an exponential speedup over classical computers for numerical rank estimation of matrices specified via sparse access (unless the one clean qubit model can be efficiently simulated on a classical computer).
Still, it is also interesting to consider settings where the matrix is specified via a different input model.
In the remainder of this section we study two examples of different input models.
Firstly, motivated by a more practical perspective we consider a seemingly weaker input model that is more closely related to the input models that appear in classical data analysis settings.
Secondly, we consider a likely stronger input model that appears throughout quantum machine learning literature, which is more informative from a complexity-theoretic perspective.

In typical (classical) applications, matrices are generally not specified via sparse access.
Here we consider an input model that is more closely related to what is encountered in a typical classical setting.
Specifically, we consider the case where a sparse matrix $A$ of size $2^n \times 2^n$ is specified as a list of triples 
\begin{align*}
\big\{(i_k,j_k,A_{i_k, j_k}) \mid A_{i_k, j_k} \neq 0 \big\},
\end{align*}
which is sorted lexicographically by column and then row. 
Storing matrices in this type of memory structure is very natural when dealing with matrices with a limited number of nonzero entries (which we denote by $\mathsf{nnz}$).
Now, for the quantum analogue we consider the same specification but we suppose that it is stored in a QRAM-type memory, only additionally allowing us to query it in superposition as follows:
\begin{align*}
  \sum_k \alpha_k \ket{k}\ket{0} \mapsto \sum_k \alpha_k \ket{k}\ket{i_k, j_k, A_{i_k, j_k}}.
\end{align*}
Since the list is sorted, and since $A$ is sparse, we can still simulate column-wise sparse access in $\bigO{\log \mathsf{nnz}}$ queries, essentially by using binary search.
Therefore, if $A$ is Hermitian, then the quantum algorithm can estimate its numerical rank in time $\bigO{\mathrm{poly}\left(n, \log  \mathsf{nnz}\right)}$.
On the other hand, the best known classical algorithms run in time $\bigO{\mathsf{nnz}}$~\cite{ubaru:approximate_rank, cheung:rank, napoli:eigenvalue_counts, lin:spectral_density}.
Consequently, the quantum algorithm achieves a speedup over the best known classical algorithm if $\mathsf{nnz}$ is at least a high-enough degree polynomial in $n$ (and it achieves an exponential speedup if $\mathsf{nnz}$ is itself exponential).
For the case where $A$ is not Hermitian, recall that we also need sparse access to $A^\dagger$.
For this issue we found no general method that can do so in time less than $\bigO{\mathsf{nnz}}$, without assuming a high sparsity.
However, the high sparsity then exactly offsets any potential quantum advantage in the full algorithm complexity. 

Next, we consider a likely stronger input model which is widely-studied in the quantum machine learning literature.
Specifically, we study the quantum-accessible data structure introduced in~\cite{kerenidis:qrs,kerenidis:qram}, which can generate quantum states proportional to the columns of the input matrix, together with a quantum state whose amplitudes are proportional to the 2-norms of the columns.
When the input matrix is provided in this quantum-accessible data structure, the quantum-algorithmic methods of~\cite{gilyen:block, chakraborty:block} can be used to estimate its numerical rank in time $\bigO{\mathrm{poly}(A_{\text{max}}, n)}$, where $A_{\text{max}} = \max_{i, j}|A_{ij}|$.

The classical analogue of this quantum-accessible data structure is the sampling and query access model introduced in~\cite{tang:dequantization}, which brought forth the ``dequantization'' methods discussed in~\cite{chia:dequantizations}.
At present it is not clear whether assuming sampling and query access allows us to efficiently estimate the numerical rank using dequantizations, or other methods.
Here both possibilities are interesting.
Firstly, if numerical rank estimation remains equally hard with sampling and query access, then it shows that quantum algorithms relying on the methods of LGZ have a chance of maintaining their exponential advantage in more general scenarios.
Secondly, if an efficient classical algorithm for numerical rank estimation is possible with sampling and query access, then this leads to new insights regarding the hardness of the one clean qubit model.
Recall that we have shown that estimating the numerical rank of matrices specified via sparse access is $\mathsf{DQC1}$-hard (in the sense that, if a classical algorithm could do so efficiently given analogous access, then it can be used to efficiently solve all problems in $\mathsf{DQC1}$).
Now for the sparse matrix case, the only difference between sparse access and sampling and query access is that the latter allows one to sample from a distribution whose probabilities are proportional to the 2-norms of the columns.
Indeed, the other part (i.e., sampling from distributions whose probabilities are proportional to the squared entries of the columns) is straightforward when the matrix is specified via sparse access.
This implies that, if sampling and query access allows us to efficiently estimate the numerical rank of sparse matrices, then producing samples according to the 2-norms of the columns of a sparse matrix is $\mathsf{DQC1}$-hard. 
This also holds for the log-local Hamiltonian setting, so it would also follow that sampling from a distribution proportional to the 2-norms of the columns of log-local Hamiltonians is $\mathsf{DQC1}$-hard.
We summarize this observation in the proposition below.

\begin{proposition}
Suppose there exists an efficient classical algorithm for numerical rank estimation (or, equivalently \textsc{llsd}) for matrices provided by sampling and query access.
Then, sampling from a distribution whose probabilities are proportional to the 2-norms of the columns of a sparse Hermitian matrix is $\mathsf{DQC1}$-hard (in the sense that, if a classical algorithm could do so efficiently, then it can efficiently simulate the one clean qubit model).
\end{proposition}

\subsection{Combinatorial Laplacians beyond Betti numbers
  \label{subsec:comb_lapl_beyond_betti}}

In the previous section we discussed a practical application of the quantum-algorithmic methods behind the algorithm for Betti number estimation by using the same methods, but changing the family of input matrices (i.e., going beyond combinatorial Laplacians).
In this section we take a different approach, namely we again consider the combinatorial Laplacians, but investigate applications beyond Betti number estimation (i.e., beyond estimating its nullity) relying on different algorithms than the one for low-lying spectral density estimation.
Moreover, we will again find regimes where the same type of evidence of classical hardness can be provided, further motivating investigations into quantum algorithms that operate on the combinatorial Laplacians.

The eigenvalues and eigenvectors of the combinatorial Laplacian have many interesting graph-oriented applications beyond the applications in topological data analysis discussed in Section~\ref{sec:tda}.
The intuition behind this is that the combinatorial Laplacian can be viewed as a generalization of the standard graph Laplacian.
For example, there exist generalizations of spectral clustering and label propagation (important techniques in machine learning that are used for dimensionality reduction and classification) which utilize the eigenvalues and eigenvectors of the combinatorial Laplacians~\cite{osting:simplicial_clustering_label_propagation}.
Moreover, the eigenvalues of a normalized version of the combinatorial Laplacian convey information about the existence of circuits of cliques (i.e., ordered lists of adjacent cliques that cover the whole graph) and about the chromatic number~\cite{horak:spectra_comb_laplacian}.
Lastly, Kirchhoff's matrix tree theorem -- which relates the eigenvalues of the standard graph Laplacian to the number of spanning trees -- turns out to have a generalization to higher-order combinatorial Laplacians~\cite{duval:matrix-tree}.

The specific problem that we study in this section is that of sampling from a distribution over the eigenvalues whose probabilities are proportional to the magnitude of the eigenvalues.
In particular, we give a quantum algorithm that efficiently samples from an approximation of these distributions.
Moreover, we show that sampling from these distributions for arbitrary sparse Hermitian matrices is again as hard as simulating the one clean qubit model, which shows that it is classically intractable (unless the one clean qubit model can be efficiently simulated on a classical computer).
Finally, we discuss how this quantum algorithm can speed up spectral entropy estimation, which when applied to combinatorial Laplacians can be used to compare complex networks.

We define the problem that we study in this section as follows.

\vspace{5pt}
\noindent\begin{tabularx}{\linewidth}{l X c}
  \multicolumn{2}{l}{\textbf{\textsf{Sparse weighted eigenvalue sampling (SWES)}}} \\
  \multicolumn{2}{l}{\textbf{Input:}}\\
  1) & A sparse positive semidefinite matrix $H \in \mathbb{C}^{2^n \times 2^n}$, with $||H|| \leq 1$  and $\tr{H}/2^n \in \bigO{\mathrm{poly}(n)}$.\\
  2) & An estimation precision $\delta \in \Omega\left(1/\mathrm{poly}(n)\right)$.\\
  3) & A sampling error probability $\mu \in \Omega\left( 1/\mathrm{poly}(n)\right)$. \\
  \end{tabularx}
\noindent\begin{tabularx}{\linewidth}{l X c}
  \textbf{Output:} & A sample drawn from a $(\delta, \mu)$-approximation of the distribution $p(\lambda_j) = \lambda_j / \tr{H}$.
\end{tabularx}
\vspace{5pt}

Using the subroutines of the quantum algorithm for Betti number estimation (i.e., Hamiltonian simulation and quantum phase estimation), we can efficiently sample from an approximation of the distribution of \textsc{swes} defined above.
In fact, we can efficiently implement \emph{purified quantum query-access} to $p(\lambda_j)$~\cite{gilyen:entropy}.
To be precise, we can implement an approximation of the unitary $U_H$ (and its inverse) which acts as
\begin{align}
U_H \ket{0}_A\ket{0}_B = \ket{\psi_H} = \sum_{j = 0}^{2^n-1}\sqrt{p\left( \lambda_j\right)}\ket{\psi_j}_A\ket{\phi_j}_B,
\label{eq:quantum_access}
\end{align}
such that $\Tr_B\left(\ket{\psi_H}\bra{\psi_H}\right) = H / \tr{H}$.
Purified quantum query-access has been shown to be more powerful than standard classical sampling access, as it can speedup the postprocessing of the samples when trying to find out properties of the underlying distribution~\cite{gilyen:entropy}.

We implement an approximation of the purified quantum-query access defined in Eq.~\eqref{eq:quantum_access} as follows:
\begin{enumerate}
\item Prepare the following input state by taking a maximally entangled state (which can always be expressed in the eigenbasis of $H$ in one of its subsystems) and adding two ancillary registers
  \begin{align*}
    \ket{\psi}_{in} = \frac{1}{\sqrt{2^n}} \sum_{k = 0}^{2^n-1}\ket{\psi_k}\ket{\phi_k} \otimes \ket{0^t} \otimes \ket{0}_{flag},
  \end{align*}
  where $\{\ket{\psi_k}\}_{k=0}^{2^n-1}$ are orthonormal eigenvectors of $H$ and $\{\ket{\phi_k}\}_{k=0}^{2^n-1}$ is an orthonormal basis of~$\mathbb{C}^{2^n}$.
\item Use Hamiltonian simulation on $H$, and apply quantum phase estimation of the realized unitary to the first register to prepare the state
  \begin{align*}
    \frac{1}{\sqrt{2^n}}&\sum_{k = 0}^{2^n-1}\sum_{j = 0}^{2^t}\alpha_{k,j}\ket{\psi_k}\ket{\phi_k}\otimes \ket{\widetilde{\lambda_{k,j}}} \otimes \ket{0}_{flag} \\
    &\approx \frac{1}{\sqrt{N}}\sum_{k = 0}^{2^n-1}\ket{\psi_k}\ket{\phi_k}\otimes \ket{\widetilde{\lambda_{k}}} \otimes \ket{0}_{flag},
  \end{align*}
  where the $\widetilde{\lambda_{k,j}}$ are $t$-bit strings, $|\alpha_{k,j}|^2$ is close to 1 if and only if $\lambda_k \approx \widetilde{\lambda_{k,j}}$, and $\widetilde{\lambda_k}$ denotes the best $t$-bit approximation of $\lambda_k$.
\item Use controlled rotations to ``imprint'' the $t$-bit approximations of the eigenvalues into the amplitudes of the flag-register to prepare the state
  \begin{align*}
    &\frac{1}{\sqrt{2^n}}\sum_{k = 0}^{2^n-1}\sum_{j = 0}^{2^t}\alpha_{k,j}\ket{\psi_k}\ket{\phi_k}\otimes \ket{\widetilde{\lambda_{k,j}}} \\
   &\quad\quad\quad \otimes \left(\sqrt{\widetilde{\lambda_{k,j}}}\ket{0}_{flag} + \sqrt{1 - \widetilde{\lambda_{k,j}}}\ket{1}_{flag}\right) \\
    &\approx \frac{1}{\sqrt{2^n}}\sum_{k = 0}^{2^n-1}\ket{\psi_k}\ket{\phi_k}\otimes \ket{\widetilde{\lambda_{k}}} \\
    &\quad\quad\quad\otimes \left(\sqrt{\widetilde{\lambda_{k}}}\ket{0}_{flag} + \sqrt{1 - \widetilde{\lambda_{k}}}\ket{1}_{flag}\right).
  \end{align*}
\item Use fixed point amplitude amplification to amplify states whose flag-register is in the state~$\ket{0}$ to prepare an approximation of the state
  \begin{align*}
    \qquad\frac{1}{\sqrt{\Tr(H)}}&\sum_{k = 0}^{2^n-1}\sum_{j = 0}^{2^t}\alpha_{k,j}\sqrt{\widetilde{\lambda_{k,j}}}\ket{\psi_k}\ket{\phi_k}\otimes \ket{\widetilde{\lambda_{k,j}}} \otimes \ket{0}_{flag}\\
    &\hspace{-23pt}\approx \frac{1}{\sqrt{\Tr(H)}}\sum_{k = 0}^{2^n-1}\sqrt{\widetilde{\lambda_{k}}}\ket{\psi_k}\ket{\phi_k}\otimes \ket{\widetilde{\lambda_{k}}} \otimes \ket{0}_{flag}.
  \end{align*}
  \item Finally, uncompute and discard the eigenvalue- and flag-register to prepare the state
  \begin{align*}
    \ket{\psi_H} &= \frac{1}{\sqrt{\Tr(H)}}\sum_{k = 0}^{2^n-1}\sum_{j = 0}^{2^t}\alpha_{k,j}\sqrt{\widetilde{\lambda_{k,j}}}\ket{\psi_k}\ket{\phi_k} \\
    &\approx \frac{1}{\sqrt{\Tr(H)}}\sum_{k = 0}^{2^n-1}\sqrt{\widetilde{\lambda_{k}}}\ket{\psi_k}\ket{\phi_k}.
  \end{align*}
\end{enumerate}
Looking at the cost of the above algorithm, we note that Steps 2~and~3 can be implemented up to polynomial precision in time $\bigO{\mathrm{poly}(n)}$.
Also, note that Step~4 can be implemented up to polynomial precision in time $\bigO{\sqrt{2^n/\tr{H}}}$, which brings the total runtime to
\[
\bigO{\mathrm{poly}(n) + \sqrt{2^n/\tr{H}}}.
\]

Besides being able to efficiently sample from an approximation of \textsc{swes} on a quantum computer, we show that \textsc{swes} requires superpolynomial time on a classical computer (unless the one clean qubit model can be efficiently simulated on a classical computer). 
To be precise, we show that sampling from \textsc{swes} allows us to efficiently estimate the normalized subtrace discussed in Section~\ref{subsec:results}, which is known to be $\mathsf{DQC1}$-hard~\cite{brandao:thesis}.
We gather this in the following theorem, the proof of which can be found in the Supplementary Material.
\begin{restatable}{theorem}{sweshardness}
  \textsc{swes} is $\mathsf{DQC1}$-hard.
  Moreover, \textsc{swes} with the input restricted to log-local Hamiltonians remains $\mathsf{DQC1}$-hard.
 \label{thm:dqc1_hardness_swes}
\end{restatable}

The above theorem motivates us to look for practical applications of \textsc{swes}, or more specifically, of the purified quantum query-access described in Eq.~\eqref{eq:quantum_access}.
We end this section by discussing such an application called spectral entropy estimation, which when applied to combinatorial Laplacians can be used to compare complex networks.
The classical hardness of \textsc{swes} opens up another road towards practical quantum advantage, as it could be that combinatorial Laplacians arising in complex network analysis form a rich enough family for which \textsc{swes} remains classically hard when restricted to them.

\subsubsection{Spectral entropy estimation of the combinatorial Laplacian
\label{sucsubsec:spectral_entropy}}

Recently, several quantum information-inspired entropic measures for complex network analysis have been proposed~\cite{biamonte:entropy_complex_network_nature, domenico:entropy_complex_network_prx}.
One example of these are spectral entropies of the combinatorial Laplacian, which measure the degree of overlapping of cliques within the given complex network~\cite{passerini:entropy_complex_network, simmons:entropy_complex_network, maletic:entropy_complex_network}.
Specifically, it has been shown that these entropic measures can be used to measure network centralization (i.e., how central is the most central node in relation to all other nodes)~\cite{simmons:entropy_complex_network}, network regularity (i.e., the difference in degrees among nodes)~\cite{passerini:entropy_complex_network}, and clique connectivity (i.e., the overlaps between communities in the network)~\cite{maletic:entropy_complex_network}. 

If $\lambda_0, \dots, \lambda_{d_k^G-1}$ denote the eigenvalues of a combinatorial Laplacian $\Delta_k^G$ (i.e., $d^G_k = \dim \mathcal{H}_k^G)$, then its \emph{spectral entropy} is defined by
\begin{align}
  \label{eq:spectral_entropy}
  S(\Delta_k^G) = -\sum_{j = 0}^{d^G_k-1}p(\lambda_j)\log(p(\lambda_j)),
\end{align}
where we define $p(\lambda_j) = \lambda_j / \left(\sum_k\lambda_k\right)$.
This spectral entropy coincides with the von Neumann entropy of $\Delta_k^G/ \tr{\Delta_k^G}$.
Equivalently, it coincides with the Shannon entropy of the distribution~$p(\lambda_j)$.
Another entropy that is used in complex network analysis is the \emph{$\alpha$-Renyi spectral entropy}, which is given by
\begin{align}
  \label{eq:renyi_entropy}
  S_\alpha(\Delta_k^G) = \frac{1}{1 - \alpha}\log\left(\sum_{j = 0}^{d_k-1}p(\lambda_j)^\alpha\right),
\end{align}
where $\alpha \geq 0$ and $\alpha \neq 1$.
The limit for $\alpha \rightarrow 1$ is the spectral entropy as defined in Eq.~\eqref{eq:spectral_entropy}.

To estimate the spectral entropy defined in Eq.~\eqref{eq:spectral_entropy}, one can use techniques from~\cite{acharya:entropy, valiant:entropy} to classically postprocess samples from $p(\lambda_j)$ that one obtains from the quantum algorithm for \textsc{swes} described in the previous section.
However, since we can implement purified quantum query-access using the algorithm described in the previous section, the postprocessing can be sped up quadratically using quantum methods~\cite{gilyen:entropy}.
This idea of speeding up the postprocessing of samples using quantum methods also holds for the $\alpha$-Renyi entropy defined in Eq.~\eqref{eq:renyi_entropy}, where one can either classically postprocess the samples~\cite{acharya:renyi_entropy}, or use faster quantum methods~\cite{subramanian:entropy}.

Because we have shown that sampling from \textsc{swes} is $\mathsf{DQC1}$-hard, the above approach to spectral entropy estimation can not be done efficiently on a classical computer -- i.e., it cannot be dequantized -- when generalized to arbitrary sparse matrices (unless the one clean qubit model can be efficiently simulated on a classical computer).
Moreover, as the $\alpha$-Renyi entropy is the logarithm of the Schatten $p$-norm, and it is known that estimating Schatten $p$-norms is $\mathsf{DQC1}$-hard~\cite{cade:schatten_p}, we find that computing $\alpha$-Renyi entropy is classically intractable (again, unless the one clean qubit model can be efficiently simulated on a classical computer). 

\section{Possibilities and challenges for implementations
\label{sec:nisq}}

As near-term quantum devices are still limited, it is crucial to make sure to use them to their fullest extent when implementing a quantum algorithm.
Near-term devices are limited in size, gates are error prone, qubits decohere, and their architectures are limited~\cite{preskill:nisq}.
We are therefore interested in algorithms that require few gates (to minimize the effect of decoherence and gate errors), that are not too demanding regarding architecture, while achieving advantages with few qubits and being tolerant to noise (which will inevitably be present in the system regardless of the depth and gate count). 
The quantum algorithms we consider use Hamiltonian simulation and quantum phase estimation.
Fortunately, both resource optimization~\cite{bauer:quantum_simulation} and error-mitigation~\cite{temme:error, bonet:error, endo:error, mcardle:error, tom:error} for these routines are important topics for the broadly investigated field of quantum algorithms for quantum chemistry and many-body physics, and any progress achieved for those purposes can be readily applied.
Moreover, recent work has focused on reducing the depth of the quantum circuit required to implement the algorithm for (approximate) Betti number estimation~\cite{ubaru:qtda}.
In this section we will focus on the issues of size and noise.
First, we investigate the required number of qubits and we propose methods on how to reduce this.
Based on these methods, we provide an estimate of the number of qubits required to challenge classical methods.
Finally, we discuss issues regarding robustness of the algorithm to noise in the quantum hardware.

To analyze the number of qubits required to implement Hamiltonian simulation of a $2^n \times 2^n$-sized input matrix, we consider two possible scenarios:\ the input matrix is either given to us as local terms, or it is specified via sparse access.
If the input matrix is given to us as local terms, then we can implement Hamiltonian simulation based on the Trotter-Suzuki formula~\cite{lloyd:hs}.
As this Hamiltonian simulation technique does not require ancillary qubits (assuming the available gate set can implement each of the Trotter steps without ancillary qubits)~\cite{cade:schatten_p}, we can implement it using only $n$ qubits.
On the other hand, if the input matrix is specified via sparse access, then we have to use more intricate Hamiltonian simulation techniques (e.g., based on quantum signal processing~\cite{low:hs}).
The downside of these methods is that they require an ancillary register to `load' the queries to the sparse-access oracles onto.
By having to add this ancillary register, the total number of qubits required to implement these Hamiltonian simulation techniques becomes $2n + r + 1$, where $r$ is the number of bits used to specify the entries of the input matrix.
In other words, sparse-access oracles more than double the required number of qubits.

When possible it is therefore advantageous to avoid using sparse access when having first proof-of-principle demonstrations of quantum advantage in mind.
One way of doing so is to add an extra precompilation step that finds a suitable decomposition of the input matrix.
In particular, one can trade-off the required number of ancilla qubits for some amount of precompilation and some extra depth of the precompiled circuit, in the following two ways.
First, one could decompose the input matrix in terms of a linear combination of unitaries, and use related techniques for Hamiltonian simulation of such input matrices~\cite{berry:hs}.
This brings the required number of qubits down from $2n + r + 1$ to $n + \log(m)$, where $m$ is the number of terms in the linear combination of unitaries.
Secondly, one could decompose the input matrix in terms of a sum of local Hamiltonians and use Hamiltonian simulation based on the Trotter-Suzuki formula.
This brings the required number of qubits down from $2n + r + 1$ to~$n$.
Thus, both approaches can halve the number of required qubits,
however, one has to be careful as finding such decompositions may constitute a dominating overhead.

In case of Betti number estimation, we note that such precompilation is in fact feasible and meaningful.
This is due to the fact that in this case there is a direct way to decompose input matrix (i.e., the combinatorial Laplacian) as a sum of Pauli-strings in order to implement Hamiltonian simulation based on the Trotter-Suzuki formula.
Specifically, due to the close relationship between combinatorial Laplacians and Hamiltonians of the fermion hardcore model (as described in Section~\ref{subsubsec:discussion})~\cite{cade:complexity} we can decompose the combinatorial Laplacian into a sum of Pauli-strings by applying a fermion to qubit mapping such as the Jordan-Wigner or Bravyi-Kitaev transformations to Eq~\eqref{eq:ham_fermion}. 
Note however that this does not guarantee that Hamiltonian simulation based on the Trotter-Suzuki formula will be efficient as the decomposition might require exponentially many terms and the locality of the individual terms could be large.
As can be seen in Eq.~\eqref{eq:ham_fermion}, the number of terms in the decomposition scales with the degree of the vertices in the complement of the graph.
In particular, if the graph is such that any vertex is connected to all other vertices except for a constant number of them, then the number of terms in the decomposition scales polynomially.
As discussed in Section~\ref{subsec:graphs_quantum_advantage}, these are exactly the type of graphs where the quantum algorithm for Betti number estimation achieves a speedup over the best known classical algorithms, since these types of graphs are clique-dense (i.e., they satisfy Eq.~\eqref{eq:clique_dense}).
The locality of the Pauli-strings in the decomposition can however not be guaranteed to be small, but this fortunately has less effect on the depth of the circuit.
Finally, we remark that this decomposition also gives rise to a technique that allows one to control the depth of the circuit required for the Hamiltonian simulation.
Namely, by dropping certain terms from the decomposition (e.g., terms with a small coefficient) one could reduce the depth of the circuit required for Hamiltonian simulation, while making sure to not perturb the matrix too much as to drastically change the low-lying spectral density.

Next, we focus on the number of qubits required for the quantum phase estimation.
Standard quantum phase estimation requires an eigenvalue register of $t$ qubits to estimate the eigenvalues up to $t$-bits of precision (which consequently determines the threshold in low-lying spectral density estimation). 
Fortunately, much improvement is possible in terms of the size of this eigenvalue register.
First, as low-lying spectral density is only concerned with whether the $t$-bit approximation of an eigenvalue is zero or not, we can bring the size the of eigenvalue register down to $\log(t)$ by using a counter~\cite{rennela:counter}.
Moreover, we can bring the size of this eigenvalue register down to a single qubit at the expense of classical post-processing and qubit reinitialization methods~\cite{dutkiewicz:qpe, tom:qpe, somma:qpe}.

We can now give the brief estimate of the number of qubits needed for demonstrations of quantum advantage (i.e., sizes needed to go beyond the best known classical methods).
The best known classical methods for low-lying spectral density estimation, to our knowledge, are able to estimate the rank of a matrix in time linear in the number of nonzero entries~\cite{ubaru:approximate_rank, cheung:rank, napoli:eigenvalue_counts, lin:spectral_density}.
These methods are at most quadratically faster than exact diagonalization, which tends to hit a practical wall around matrices of size $2^{40}$.
We therefore look at how many qubits are required to estimate the low-lying spectral density below a threshold of about $10^{-9}$ (i.e., $t \approx \log(10^{9}) < 30$) of matrices of size around $2^{80}$ (i.e., $n \approx 80$).
In this case, the required number of qubits for standard implementations is approximately
\[
  2n + r + 1 + t \approx 200.
\]
If we precompile the input matrix through finding a decomposition in terms of local Hamiltonians, this can be reduced to
\[
n + t \approx 110.
\]
This can be further reduced to $n + \log(t)$ by using a counter in the eigenvalue register.
Lastly, by using a single-qubit eigenvalue register (at the cost of classical postprocessing and qubit reinitialization) we bring the number of required qubits in the optimal case down to 
\[
n + 1 \approx 80,
\]
which is tantalizingly close to what leading teams are expected to achieve in the immediate future in terms of qubit numbers alone.

When it comes to the robustness to noise in the hardware, we need to consider the type of algorithm that is being applied (i.e., how noise affects this algorithm in general) together with the specifics of the application.
The algorithm we consider involves many iterations of Hamiltonian simulation and quantum phase estimation, where we are interested in the expected value of a two outcome measurement (designating the zero eigenvalues).
As noted earlier, these routines are also crucial for quantum algorithms for quantum chemistry and many-body physics, and consequently, all error-mitigation methods developed for these purposes can be readily applied~\cite{temme:error, bonet:error, endo:error, mcardle:error, tom:error}.
However, as in quantum chemistry and many-body physics one extracts the entire eigenvalues, as opposed to just the frequency of the zero eigenvalue, the application we consider is less demanding.
Additional robustness properties van be inferred from the nature of the particular problem solved.
For instance, in machine learning and data analysis applications, the fact that the algorithm serves the purpose of dealing with noise in the data might make noise in the hardware less detrimental compared to when solving more exact problems~\cite{dunjko:review}.

Unfortunately, this argument cannot be as readily applied to Betti number estimation, as noise in the data does not correspond to small perturbations of the simulated matrix (i.e., the combinatorial Laplacian), but rather to a completely different matrix altogether.
In turn, small perturbations of the simulated matrix do not corresponds to any meaningful perturbation of the input data.
However, we can still identify certain robust features by considering what perturbations of the combinatorial Laplacian entail for the final output, i.e., the low-lying spectral density.
Specifically, if the combinatorial Laplacian is perturbed by a small enough matrix (e.g., in terms of operator norm or rank), then the low-lying spectral density remains largely unchanged as such perturbations will not push the low-lying eigenvalues above the threshold.
These settings are often studied in the field of perturbation theory~\cite{kato:perturbation}, which would allow us to make these arguments completely formal.
Moreover, as a random matrix is likely of full rank~\cite{feng:rank}, the perturbed combinatorial Laplacian is also likely of full rank, indicating that in the noisy setting we should focus on approximate Betti number estimation methods, as opposed to exact ones.
Finally, there has been work verifying the robustness of the quantum algorithm for Betti number estimation in an experimental setting~\cite{huang:demonstration}.

\section{Summary
  \label{sec:summary}}

In this paper we investigated the potential of a class of problems arising from the quantum algorithm for topological data analysis~\cite{lloyd:lgz_algorithm} to become genuinely useful applications of unrestricted, or even near-term, quantum computers with a superpolynomial quantum speedup. 
We showed that this algorithm along with a number of new algorithms provided by us (with applications in numerical linear algebra, machine learning and complex network analysis) solve problems that are classically intractable under widely-believed complexity-theoretic assumptions by showing that they are as hard as simulating the one clean qubit model.
While the complete resolution of the hardness of the topological data analysis problem will require future research into the properties of the combinatorial Laplacians (which as we showed is also interesting for other applications such as complex network analysis), our results eliminate the possibility of generic dequantization methods that are oblivious to the structure of the combinatorial Laplacian.
Specifically, our results showed that the methods of the quantum algorithm for topological data analysis withstand the sweeping dequantization results of Tang et al.~\cite{tang:dequantization, chia:dequantizations}.
To  analyze whether  it  is  possible  to  further  strengthen  the  argument for quantum advantage (or, to actually find an efficient classical algorithm) for the narrow TDA problem, we investigated state-of-the-art  classical algorithms and  we highlighted the theoretical hurdles that, at least currently, stymie such classical approaches.
Regarding near-term implementations, we identified that implementing sparse access to the input matrix is a major bottleneck in terms of the required number of qubits, we proposed multiple methods to circumvent this bottleneck via classical precompilation strategies, and we investigated the required resources to challenge the best known classical methods.
In summary, our results show that the quantum-algorithmic methods behind the algorithm for topological data analysis give rise to a source of both useful and guaranteed superpolynomial quantum speedups (that are amenable to near-term restricted quantum computers), recovering some of the potential for linear-algebraic quantum machine learning to become one of quantum computing's killer applications.

\begin{acknowledgments}
The authors thank Dorit Aharonov, Tomoyuki Morimae and Ronald de Wolf for early discussions on the topic.
The authors are grateful to Ronald de Wolf for careful reading of the manuscript and for giving valuable comments.

This work was supported by the Dutch Research Council (NWO/OCW), as part of the Quantum Software Consortium programme (project number 024.003.037), and through QuantERA project QuantAlgo 680-91-034.
\end{acknowledgments}

\bibliographystyle{quantum}
\bibliography{quantum_advantage_tda}

\begin{thebibliography}{10}

\bibitem{dunjko:review}
Vedran Dunjko and Peter Wittek.
\newblock ``A non-review of quantum machine learning: trends and
  explorations''.
\newblock
  \href{https://dx.doi.org/https://doi.org/10.22331/qv-2020-03-17-32}{Quantum
  {\bf 4}, 32}~(2020).

\bibitem{biamonte:review}
Jacob Biamonte, Peter Wittek, Nicola Pancotti, Patrick Rebentrost, Nathan
  Wiebe, and Seth Lloyd.
\newblock ``Quantum machine learning''.
\newblock \href{https://dx.doi.org/https://doi.org/10.1038/nature23474}{Nature
  {\bf 549}, 195--202}~(2017).
\newblock  \href{http://arxiv.org/abs/1611.09347}{arXiv:1611.09347}.

\bibitem{harrow:hhl}
Aram~W Harrow, Avinatan Hassidim, and Seth Lloyd.
\newblock ``Quantum algorithm for linear systems of equations''.
\newblock
  \href{https://dx.doi.org/https://doi.org/10.1103/PhysRevLett.103.150502}{Physical
  review letters {\bf 103}, 150502}~(2009).
\newblock  \href{http://arxiv.org/abs/0811.3171}{arXiv:0811.3171}.

\bibitem{havlivcek:pqcs}
Vojt{\v{e}}ch Havl{\'\i}{\v{c}}ek, Antonio~D C{\'o}rcoles, Kristan Temme,
  Aram~W Harrow, Abhinav Kandala, Jerry~M Chow, and Jay~M Gambetta.
\newblock ``Supervised learning with quantum-enhanced feature spaces''.
\newblock
  \href{https://dx.doi.org/https://doi.org/10.1038/s41586-019-0980-2}{Nature
  {\bf 567}, 209--212}~(2019).
\newblock  \href{http://arxiv.org/abs/1804.11326}{arXiv:1804.11326}.

\bibitem{schuld:pqcs}
Maria Schuld, Alex Bocharov, Krysta~M Svore, and Nathan Wiebe.
\newblock ``Circuit-centric quantum classifiers''.
\newblock
  \href{https://dx.doi.org/https://doi.org/10.1103/PhysRevA.101.032308}{Physical
  Review A {\bf 101}, 032308}~(2020).
\newblock  \href{http://arxiv.org/abs/1804.00633}{arXiv:1804.00633}.

\bibitem{benedetti:pqcs}
Marcello Benedetti, Erika Lloyd, Stefan Sack, and Mattia Fiorentini.
\newblock ``Parameterized quantum circuits as machine learning models''.
\newblock
  \href{https://dx.doi.org/https://doi.org/10.1088/2058-9565/ab4eb5}{Quantum
  Science and Technology {\bf 4}, 043001}~(2019).
\newblock  \href{http://arxiv.org/abs/1906.07682}{arXiv:1906.07682}.

\bibitem{tang:dequantization}
Ewin Tang.
\newblock ``A quantum-inspired classical algorithm for recommendation
  systems''.
\newblock
  \href{https://dx.doi.org/https://doi.org/10.1145/3313276.3316310}{Proceedings
  of the 51st Annual ACM SIGACT Symposium on Theory of Computing}~(2019).
\newblock  \href{http://arxiv.org/abs/1807.04271}{arXiv:1807.04271}.

\bibitem{chia:dequantizations}
Nai-Hui Chia, Andr{\'a}s Gily{\'e}n, Tongyang Li, Han-Hsuan Lin, Ewin Tang, and
  Chunhao Wang.
\newblock ``Sampling-based sublinear low-rank matrix arithmetic framework for
  dequantizing quantum machine learning''.
\newblock
  \href{https://dx.doi.org/https://doi.org/10.1145/3357713.3384314}{Proceedings
  of the 52nd Annual ACM SIGACT symposium on theory of computing}~(2020).
\newblock  \href{http://arxiv.org/abs/1910.06151}{arXiv:1910.06151}.

\bibitem{kerenidis:qrs}
Iordanis Kerenidis and Anupam Prakash.
\newblock ``Quantum recommendation systems''.
\newblock
  \href{https://dx.doi.org/http://dx.doi.org/10.4230/LIPIcs.ITCS.2017.49}{Proceedings
  of the 8th Innovations in Theoretical Computer Science Conference}~(2017).
\newblock  \href{http://arxiv.org/abs/1603.08675}{arXiv:1603.08675}.

\bibitem{lloyd:qpca}
Seth Lloyd, Masoud Mohseni, and Patrick Rebentrost.
\newblock ``Quantum principal component analysis''.
\newblock \href{https://dx.doi.org/https://doi.org/10.1038/nphys3029}{Nature
  Physics {\bf 10}, 631--633}~(2014).
\newblock  \href{http://arxiv.org/abs/1307.0401}{arXiv:1307.0401}.

\bibitem{babbush:poly_speedup}
Ryan Babbush, Jarrod McClean, Craig Gidney, Sergio Boixo, and Hartmut Neven.
\newblock ``Focus beyond quadratic speedups for error-corrected quantum
  advantage''.
\newblock
  \href{https://dx.doi.org/https://doi.org/10.1103/PRXQuantum.2.010103}{Physical
  review X Quantum}~(2021).
\newblock  \href{http://arxiv.org/abs/2011.04149}{arXiv:2011.04149}.

\bibitem{lloyd:lgz_algorithm}
Seth Lloyd, Silvano Garnerone, and Paolo Zanardi.
\newblock ``Quantum algorithms for topological and geometric analysis of
  data''.
\newblock \href{https://dx.doi.org/https://doi.org/10.1038/ncomms10138}{Nature
  communications {\bf 7}, 1--7}~(2016).
\newblock  \href{http://arxiv.org/abs/1408.3106}{arXiv:1408.3106}.

\bibitem{preskill:nisq}
John Preskill.
\newblock ``Quantum computing in the {N}{I}{S}{Q} era and beyond''.
\newblock
  \href{https://dx.doi.org/https://doi.org/10.22331/q-2018-08-06-79}{Quantum
  {\bf 2}, 79}~(2018).
\newblock  \href{http://arxiv.org/abs/1801.00862}{arXiv:1801.00862}.

\bibitem{ghrist:barcodes}
Robert Ghrist.
\newblock ``Barcodes: the persistent topology of data''.
\newblock
  \href{https://dx.doi.org/https://doi.org/10.1090/S0273-0979-07-01191-3}{Bulletin
  of the American Mathematical Society {\bf 45}, 61--75}~(2008).

\bibitem{eckmann:comb_lapl}
Beno Eckmann.
\newblock ``Harmonische {F}unktionen und {R}andwertaufgaben in einem
  {K}omplex''.
\newblock
  \href{https://dx.doi.org/https://doi.org/10.1007/BF02566245}{Commentarii
  Mathematici Helvetici {\bf 17}, 240--255}~(1944).

\bibitem{friedman:computing_betti}
Joel Friedman.
\newblock ``Computing {B}etti numbers via combinatorial {L}aplacians''.
\newblock
  \href{https://dx.doi.org/https://doi.org/10.1007/PL00009218}{Algorithmica
  {\bf 21}, 331--346}~(1998).

\bibitem{govek:figure}
Kiya~W Govek, Venkata~S Yamajala, and Pablo~G Camara.
\newblock ``Clustering-independent analysis of genomic data using spectral
  simplicial theory''.
\newblock
  \href{https://dx.doi.org/https://doi.org/10.1371/journal.pcbi.1007509}{PLoS
  computational biology}~(2019).

\bibitem{gunn:review}
Sam Gunn and Niels Kornerup.
\newblock ``Review of a quantum algorithm for {B}etti numbers''~(2019).
\newblock  \href{http://arxiv.org/abs/1906.07673}{arXiv:1906.07673}.

\bibitem{low:hs}
Guang~Hao Low and Isaac~L Chuang.
\newblock ``Optimal hamiltonian simulation by quantum signal processing''.
\newblock
  \href{https://dx.doi.org/https://doi.org/10.1103/PhysRevLett.118.010501}{Physical
  review letters {\bf 118}, 010501}~(2017).
\newblock  \href{http://arxiv.org/abs/1606.02685}{arXiv:1606.02685}.

\bibitem{goldberg:comb_lapl}
Timothy~E Goldberg.
\newblock ``Combinatorial laplacians of simplicial complexes''.
\newblock Master's thesis.
\newblock Bard College.
\newblock ~(2002).

\bibitem{n&c}
Michael~A. Nielsen and Isaac~L. Chuang.
\newblock ``Quantum computation and quantum information''.
\newblock
  \href{https://dx.doi.org/https://doi.org/10.1017/CBO9780511976667}{Cambridge
  University Press}. ~(2011).

\bibitem{gundert:cheeger}
Anna Gundert and May Szedl{\'a}ky.
\newblock ``Higher dimensional discrete {C}heeger inequalities''.
\newblock
  \href{https://dx.doi.org/https://doi.org/10.1145/2582112.2582118}{Proceedings
  of the 13th annual symposium on Computational Geometry}~(2014).
\newblock  \href{http://arxiv.org/abs/1401.2290}{arXiv:1401.2290}.

\bibitem{chen:clique}
Jianer Chen, Xiuzhen Huang, Iyad~A Kanj, and Ge~Xia.
\newblock ``Strong computational lower bounds via parameterized complexity''.
\newblock
  \href{https://dx.doi.org/https://doi.org/10.1016/j.jcss.2006.04.007}{Journal
  of Computer and System Sciences {\bf 72}, 1346--1367}~(2006).

\bibitem{brandao:thesis}
Fernando~GSL Brand{\~a}o.
\newblock ``Entanglement theory and the quantum simulation of many-body
  physics''.
\newblock
  \href{https://dx.doi.org/https://doi.org/10.48550/arXiv.0810.0026}{PhD
  thesis}.
\newblock University of London.
\newblock ~(2008).

\bibitem{wocjan:bqp_complete}
Pawel Wocjan and Shengyu Zhang.
\newblock ``Several natural {B}{Q}{P}-complete problems''~(2006).
\newblock
  \href{http://arxiv.org/abs/quant-ph/0606179}{arXiv:quant-ph/0606179}.

\bibitem{brown:dos}
Brielin Brown, Steven~T Flammia, and Norbert Schuch.
\newblock ``Computational difficulty of computing the density of states''.
\newblock
  \href{https://dx.doi.org/https://doi.org/10.1103/PhysRevLett.107.040501}{Physical
  review letters}~(2011).
\newblock  \href{http://arxiv.org/abs/1010.3060}{arXiv:1010.3060}.

\bibitem{adamaszek:betti_np}
Micha{\l} Adamaszek and Juraj Stacho.
\newblock ``Complexity of simplicial homology and independence complexes of
  chordal graphs''.
\newblock
  \href{https://dx.doi.org/https://doi.org/10.1016/j.comgeo.2016.05.003}{Computational
  Geometry {\bf 57}, 8--18}~(2016).

\bibitem{gilyen:block}
Andr{\'a}s Gily{\'e}n, Yuan Su, Guang~Hao Low, and Nathan Wiebe.
\newblock ``Quantum singular value transformation and beyond: exponential
  improvements for quantum matrix arithmetics''.
\newblock
  \href{https://dx.doi.org/https://doi.org/10.1145/3313276.3316366}{Proceedings
  of the 51st Annual ACM SIGACT Symposium on Theory of Computing}~(2019).
\newblock  \href{http://arxiv.org/abs/1806.01838}{arXiv:1806.01838}.

\bibitem{kitaev:book}
Alexei~Yu Kitaev, Alexander Shen, Mikhail~N Vyalyi, and Mikhail~N Vyalyi.
\newblock ``Classical and quantum computation''.
\newblock \href{https://dx.doi.org/http://dx.doi.org/10.1090/gsm/047}{American
  Mathematical Society}. ~(2002).

\bibitem{knill:dqc1}
Emanuel Knill and Raymond Laflamme.
\newblock ``Power of one bit of quantum information''.
\newblock
  \href{https://dx.doi.org/https://doi.org/10.1103/PhysRevLett.81.5672}{Physical
  Review Letters}~(1998).
\newblock
  \href{http://arxiv.org/abs/quant-ph/9802037}{arXiv:quant-ph/9802037}.

\bibitem{shor:dqc1}
Peter~W Shor and Stephen~P Jordan.
\newblock ``Estimating {J}ones polynomials is a complete problem for one clean
  qubit''.
\newblock \href{https://dx.doi.org/https://doi.org/10.48660/07100034}{Quantum
  Information \& Computation {\bf 8}, 681--714}~(2008).
\newblock  \href{http://arxiv.org/abs/0707.2831}{arXiv:0707.2831}.

\bibitem{tomoyuki:dqc1}
Tomoyuki Morimae.
\newblock ``Hardness of classically sampling the one-clean-qubit model with
  constant total variation distance error''.
\newblock
  \href{https://dx.doi.org/https://doi.org/10.1103/PhysRevA.96.040302}{Physical
  Review A}~(2017).
\newblock  \href{http://arxiv.org/abs/1704.03640}{arXiv:1704.03640}.

\bibitem{tomoyuki:dqc1_k}
Tomoyuki Morimae, Keisuke Fujii, and Joseph~F Fitzsimons.
\newblock ``Hardness of classically simulating the one-clean-qubit model''.
\newblock
  \href{https://dx.doi.org/https://doi.org/10.1103/PhysRevLett.112.130502}{Physical
  review letters}~(2014).
\newblock  \href{http://arxiv.org/abs/1312.2496}{arXiv:1312.2496}.

\bibitem{lloyd:hs}
Seth Lloyd.
\newblock ``Universal quantum simulators''.
\newblock
  \href{https://dx.doi.org/https://doi.org/10.1126/science.273.5278.1073}{SciencePages
  1073--1078}~(1996).

\bibitem{cade:complexity}
Chris Cade and P~Marcos Crichigno.
\newblock ``Complexity of supersymmetric systems and the cohomology
  problem''~(2021).
\newblock  \href{http://arxiv.org/abs/2107.00011}{arXiv:2107.00011}.

\bibitem{cade:schatten_p}
Chris Cade and Ashley Montanaro.
\newblock ``The quantum complexity of computing {S}chatten $ p $-norms''.
\newblock
  \href{https://dx.doi.org/https://doi.org/10.4230/LIPIcs.TQC.2018.4}{13th
  Conference on the Theory of Quantum Computation, Communication and
  Cryptography}~(2018).
\newblock  \href{http://arxiv.org/abs/1706.09279}{arXiv:1706.09279}.

\bibitem{bookatz:qma}
Adam~D Bookatz.
\newblock ``Qma-complete problems''.
\newblock
  \href{https://dx.doi.org/https://doi.org/10.26421/QIC14.5-6-1}{Quantum
  Information \& Computation {\bf 14}, 361--383}~(2014).
\newblock  \href{http://arxiv.org/abs/1212.6312}{arXiv:1212.6312}.

\bibitem{childs:bose}
Andrew~M Childs, David Gosset, and Zak Webb.
\newblock ``The {B}ose-{H}ubbard model is {Q}{M}{A}-complete''.
\newblock
  \href{https://dx.doi.org/https://doi.org/10.1007/978-3-662-43948-7_26}{International
  Colloquium on Automata, Languages, and Programming}~(2014).
\newblock  \href{http://arxiv.org/abs/1311.3297}{arXiv:1311.3297}.

\bibitem{gorman:fermi}
Bryan O'Gorman, Sandy Irani, James Whitfield, and Bill Fefferman.
\newblock ``Electronic structure in a fixed basis is qma-complete''.
\newblock
  \href{https://dx.doi.org/https://doi.org/10.1103/PRXQuantum.3.020322}{Physical
  review X Quantum}~(2021).
\newblock  \href{http://arxiv.org/abs/2103.08215}{arXiv:2103.08215}.

\bibitem{horak:spectra_comb_laplacian}
Danijela Horak and J{\"u}rgen Jost.
\newblock ``Spectra of combinatorial {L}aplace operators on simplicial
  complexes''.
\newblock
  \href{https://dx.doi.org/https://doi.org/10.1016/j.aim.2013.05.007}{Advances
  in Mathematics {\bf 244}, 303--336}~(2013).
\newblock  \href{http://arxiv.org/abs/1105.2712}{arXiv:1105.2712}.

\bibitem{wang:persistent}
Rui Wang, Duc~Duy Nguyen, and Guo-Wei Wei.
\newblock ``Persistent spectral graph''.
\newblock
  \href{https://dx.doi.org/https://doi.org/10.1002/cnm.3376}{International
  journal for numerical methods in biomedical engineering {\bf 36},
  e3376}~(2020).
\newblock  \href{http://arxiv.org/abs/1912.04135}{arXiv:1912.04135}.

\bibitem{ahmadi:tutte}
Hamed Ahmadi and Pawel Wocjan.
\newblock ``On the quantum complexity of evaluating the {T}utte polynomial''.
\newblock
  \href{https://dx.doi.org/https://doi.org/10.1142/S021821651000808X}{Journal
  of Knot Theory and its Ramifications {\bf 19}, 727--737}~(2010).

\bibitem{ubaru:approximate_rank}
Shashanka Ubaru, Yousef Saad, and Abd-Krim Seghouane.
\newblock ``Fast estimation of approximate matrix ranks using spectral
  densities''.
\newblock \href{https://dx.doi.org/https://doi.org/10.1162/NECO_a_00951}{Neural
  computation {\bf 29}, 1317--1351}~(2017).
\newblock  \href{http://arxiv.org/abs/1608.05754}{arXiv:1608.05754}.

\bibitem{cheung:rank}
Ho~Yee Cheung, Tsz~Chiu Kwok, and Lap~Chi Lau.
\newblock ``Fast matrix rank algorithms and applications''.
\newblock \href{https://dx.doi.org/https://doi.org/10.1145/2528404}{Journal of
  the ACM (JACM) {\bf 60}, 1--25}~(2013).
\newblock  \href{http://arxiv.org/abs/1203.6705}{arXiv:1203.6705}.

\bibitem{napoli:eigenvalue_counts}
Edoardo Di~Napoli, Eric Polizzi, and Yousef Saad.
\newblock ``Efficient estimation of eigenvalue counts in an interval''.
\newblock \href{https://dx.doi.org/https://doi.org/10.1002/nla.2048}{Numerical
  Linear Algebra with Applications {\bf 23}, 674--692}~(2016).
\newblock  \href{http://arxiv.org/abs/1308.4275}{arXiv:1308.4275}.

\bibitem{lin:spectral_density}
Lin Lin, Yousef Saad, and Chao Yang.
\newblock ``Approximating spectral densities of large matrices''.
\newblock \href{https://dx.doi.org/https://doi.org/10.1137/130934283}{SIAM
  review {\bf 58}, 34--65}~(2016).
\newblock  \href{http://arxiv.org/abs/1308.5467}{arXiv:1308.5467}.

\bibitem{cohen:walk}
David Cohen-Steiner, Weihao Kong, Christian Sohler, and Gregory Valiant.
\newblock ``Approximating the spectrum of a graph''.
\newblock
  \href{https://dx.doi.org/https://doi.org/10.1145/3219819.3220119}{Proceedings
  of the 24th ACM SIGKDD International Conference on Knowledge Discovery \&
  Data Mining}~(2018).
\newblock  \href{http://arxiv.org/abs/1712.01725}{arXiv:1712.01725}.

\bibitem{mukherjee:walk}
Sayan Mukherjee and John Steenbergen.
\newblock ``Random walks on simplicial complexes and harmonics''.
\newblock \href{https://dx.doi.org/https://doi.org/10.1002/rsa.20645}{Random
  structures \& algorithms {\bf 49}, 379--405}~(2016).
\newblock  \href{http://arxiv.org/abs/1310.5099}{arXiv:1310.5099}.

\bibitem{parzanchevski:walk}
Ori Parzanchevski and Ron Rosenthal.
\newblock ``Simplicial complexes: spectrum, homology and random walks''.
\newblock \href{https://dx.doi.org/https://doi.org/10.1002/rsa.20657}{Random
  Structures \& Algorithms {\bf 50}, 225--261}~(2017).
\newblock  \href{http://arxiv.org/abs/1211.6775}{arXiv:1211.6775}.

\bibitem{reiher:clique}
Christian Reiher.
\newblock ``The clique density theorem''.
\newblock
  \href{https://dx.doi.org/http://dx.doi.org/10.4007/annals.2016.184.3.1}{Annals
  of Mathematics}~(2016).
\newblock  \href{http://arxiv.org/abs/1212.2454}{arXiv:1212.2454}.

\bibitem{moon:clique}
J.W. Moon and Moser L.
\newblock ``On a problem of turan''.
\newblock Publ. Math. Inst. Hung. Acad. Sci.~(1962).

\bibitem{lovasz:clique}
L{\'a}szl{\'o} Lov{\'a}sz et~al.
\newblock ``Very large graphs''.
\newblock
  \href{https://dx.doi.org/https://dx.doi.org/10.4310/CDM.2008.v2008.n1.a2}{Current
  Developments in Mathematics {\bf 2008}, 67--128}~(2009).
\newblock  \href{http://arxiv.org/abs/0902.0132}{arXiv:0902.0132}.

\bibitem{ugander:clique}
Johan Ugander, Lars Backstrom, and Jon Kleinberg.
\newblock ``Subgraph frequencies: Mapping the empirical and extremal geography
  of large graph collections''.
\newblock
  \href{https://dx.doi.org/https://doi.org/10.1145/2488388.2488502}{Proceedings
  of the 22nd international conference on World Wide Web}~(2013).
\newblock  \href{http://arxiv.org/abs/1304.1548}{arXiv:1304.1548}.

\bibitem{eden:clique}
Talya Eden, Dana Ron, and Will Rosenbaum.
\newblock ``{Almost Optimal Bounds for Sublinear-Time Sampling of k-Cliques in
  Bounded Arboricity Graphs}''.
\newblock \href{https://dx.doi.org/10.4230/LIPIcs.ICALP.2022.56}{49th
  International Colloquium on Automata, Languages, and Programming --
  ICALP}~(2022).
\newblock  \href{http://arxiv.org/abs/2012.04090}{arXiv:2012.04090}.

\bibitem{jolliffe:pca}
Ian~T Jolliffe.
\newblock ``Principal components in regression analysis''.
\newblock
  \href{https://dx.doi.org/https://doi.org/10.1007/978-1-4757-1904-8_8}{Pages
  129--155}.
\newblock Springer. ~(1986).

\bibitem{halko:random_low_rank}
Nathan Halko, Per-Gunnar Martinsson, and Joel~A Tropp.
\newblock ``Finding structure with randomness: Probabilistic algorithms for
  constructing approximate matrix decompositions''.
\newblock \href{https://dx.doi.org/https://doi.org/10.1137/090771806}{SIAM
  review {\bf 53}, 217--288}~(2011).
\newblock  \href{http://arxiv.org/abs/0909.4061}{arXiv:0909.4061}.

\bibitem{kerenidis:qram}
Iordanis Kerenidis and Anupam Prakash.
\newblock ``Quantum gradient descent for linear systems and least squares''.
\newblock
  \href{https://dx.doi.org/https://doi.org/10.1103/PhysRevA.101.022316}{Physical
  Review A {\bf 101}, 022316}~(2020).
\newblock  \href{http://arxiv.org/abs/1704.04992}{arXiv:1704.04992}.

\bibitem{chakraborty:block}
Shantanav Chakraborty, Andr{\'a}s Gily{\'e}n, and Stacey Jeffery.
\newblock ``The power of block-encoded matrix powers: Improved regression
  techniques via faster hamiltonian simulation''.
\newblock
  \href{https://dx.doi.org/https://doi.org/10.4230/LIPIcs.ICALP.2019.33}{46th
  International Colloquium on Automata, Languages, and Programming (ICALP
  2019)}~(2019).
\newblock  \href{http://arxiv.org/abs/1804.01973}{arXiv:1804.01973}.

\bibitem{osting:simplicial_clustering_label_propagation}
Braxton Osting, Sourabh Palande, and Bei Wang.
\newblock ``Spectral sparsification of simplicial complexes for clustering and
  label propagation''.
\newblock
  \href{https://dx.doi.org/https://doi.org/10.20382/jocg.v11i1a8}{Journal of
  Computational Geometry}~(2017).
\newblock  \href{http://arxiv.org/abs/1708.08436}{arXiv:1708.08436}.

\bibitem{duval:matrix-tree}
Art Duval, Caroline Klivans, and Jeremy Martin.
\newblock ``Simplicial matrix-tree theorems''.
\newblock
  \href{https://dx.doi.org/https://doi.org/10.1090/S0002-9947-09-04898-3}{Transactions
  of the American Mathematical Society {\bf 361}, 6073--6114}~(2009).
\newblock  \href{http://arxiv.org/abs/0802.2576}{arXiv:0802.2576}.

\bibitem{gilyen:entropy}
Andr{\'a}s Gily{\'e}n and Tongyang Li.
\newblock ``Distributional property testing in a quantum world''.
\newblock
  \href{https://dx.doi.org/https://doi.org/10.4230/LIPIcs.ITCS.2020.25}{11th
  Innovations in Theoretical Computer Science Conference (ITCS 2020)}~(2020).
\newblock  \href{http://arxiv.org/abs/1902.00814}{arXiv:1902.00814}.

\bibitem{biamonte:entropy_complex_network_nature}
Jacob Biamonte, Mauro Faccin, and Manlio De~Domenico.
\newblock ``Complex networks from classical to quantum''.
\newblock
  \href{https://dx.doi.org/https://doi.org/10.1038/s42005-019-0152-6}{Communications
  Physics {\bf 2}, 1--10}~(2019).
\newblock  \href{http://arxiv.org/abs/1702.08459}{arXiv:1702.08459}.

\bibitem{domenico:entropy_complex_network_prx}
Manlio De~Domenico and Jacob Biamonte.
\newblock ``Spectral entropies as information-theoretic tools for complex
  network comparison''.
\newblock
  \href{https://dx.doi.org/https://doi.org/10.1103/PhysRevX.6.041062}{Physical
  Review X{\bf 6}}~(2016).
\newblock  \href{http://arxiv.org/abs/1609.01214}{arXiv:1609.01214}.

\bibitem{passerini:entropy_complex_network}
Filippo Passerini and Simone Severini.
\newblock ``Quantifying complexity in networks: the von {N}eumann entropy''.
\newblock
  \href{https://dx.doi.org/https://doi.org/10.4018/jats.2009071005}{International
  Journal of Agent Technologies and Systems (IJATS) {\bf 1}, 58--67}~(2009).
\newblock  \href{http://arxiv.org/abs/0812.2597}{arXiv:0812.2597}.

\bibitem{simmons:entropy_complex_network}
David Simmons, Justin Coon, and Animesh Datta.
\newblock ``The quantum {T}heil index: characterizing graph centralization
  using von {N}eumann entropy''.
\newblock
  \href{https://dx.doi.org/https://doi.org/10.1093/COMNET/CNX061}{Journal of
  Complex Networks {\bf 6}, 859--876}~(2018).
\newblock  \href{http://arxiv.org/abs/1707.07906}{arXiv:1707.07906}.

\bibitem{maletic:entropy_complex_network}
Slobodan Maleti{\'c} and Milan Rajkovi{\'c}.
\newblock ``Combinatorial {L}aplacian and entropy of simplicial complexes
  associated with complex networks''.
\newblock
  \href{https://dx.doi.org/https://doi.org/10.1140/epjst/e2012-01655-6}{The
  European Physical Journal Special Topics {\bf 212}, 77--97}~(2012).

\bibitem{acharya:entropy}
Jayadev Acharya, Ibrahim Issa, Nirmal~V Shende, and Aaron~B Wagner.
\newblock ``Measuring quantum entropy''.
\newblock
  \href{https://dx.doi.org/https://doi.org/10.1109/ISIT.2019.8849572}{2019 IEEE
  International Symposium on Information Theory (ISIT)}~(2019).
\newblock  \href{http://arxiv.org/abs/1711.00814}{arXiv:1711.00814}.

\bibitem{valiant:entropy}
Gregory Valiant and Paul Valiant.
\newblock ``Estimating the unseen: an n/log(n)-sample estimator for entropy and
  support size, shown optimal via new clts''.
\newblock
  \href{https://dx.doi.org/https://doi.org/10.1145/1993636.1993727}{Proceedings
  of the forty-third annual ACM symposium on Theory of computing}~(2011).

\bibitem{acharya:renyi_entropy}
Jayadev Acharya, Alon Orlitsky, Ananda~Theertha Suresh, and Himanshu Tyagi.
\newblock ``Estimating r{\'e}nyi entropy of discrete distributions''.
\newblock
  \href{https://dx.doi.org/https://doi.org/10.1109/TIT.2016.2620435}{IEEE
  Transactions on Information Theory {\bf 63}, 38--56}~(2016).
\newblock  \href{http://arxiv.org/abs/1408.1000}{arXiv:1408.1000}.

\bibitem{subramanian:entropy}
Sathyawageeswar Subramanian and Min-Hsiu Hsieh.
\newblock ``Quantum algorithm for estimating renyi entropies of quantum
  states''.
\newblock
  \href{https://dx.doi.org/https://doi.org/10.1103/PhysRevA.104.022428}{Physical
  review A}~(2021).
\newblock  \href{http://arxiv.org/abs/1908.05251}{arXiv:1908.05251}.

\bibitem{bauer:quantum_simulation}
Bela Bauer, Sergey Bravyi, Mario Motta, and Garnet~Kin Chan.
\newblock ``Quantum algorithms for quantum chemistry and quantum materials
  science''.
\newblock
  \href{https://dx.doi.org/https://doi.org/10.1021/acs.chemrev.9b00829}{Chemical
  Reviews {\bf 120}, 12685--12717}~(2020).
\newblock  \href{http://arxiv.org/abs/2001.03685}{arXiv:2001.03685}.

\bibitem{temme:error}
Kristan Temme, Sergey Bravyi, and Jay~M Gambetta.
\newblock ``Error mitigation for short-depth quantum circuits''.
\newblock
  \href{https://dx.doi.org/https://doi.org/10.1103/PhysRevLett.119.180509}{Physical
  review letters}~(2017).
\newblock  \href{http://arxiv.org/abs/1612.02058}{arXiv:1612.02058}.

\bibitem{bonet:error}
Xavi Bonet-Monroig, Ramiro Sagastizabal, M~Singh, and TE~O'Brien.
\newblock ``Low-cost error mitigation by symmetry verification''.
\newblock
  \href{https://dx.doi.org/https://doi.org/10.1103/PhysRevA.98.062339}{Physical
  Review A}~(2018).
\newblock  \href{http://arxiv.org/abs/1807.10050}{arXiv:1807.10050}.

\bibitem{endo:error}
Suguru Endo, Simon~C Benjamin, and Ying Li.
\newblock ``Practical quantum error mitigation for near-future applications''.
\newblock
  \href{https://dx.doi.org/https://doi.org/10.1103/PhysRevX.8.031027}{Physical
  Review X}~(2018).
\newblock  \href{http://arxiv.org/abs/1712.09271}{arXiv:1712.09271}.

\bibitem{mcardle:error}
Sam McArdle, Xiao Yuan, and Simon Benjamin.
\newblock ``Error-mitigated digital quantum simulation''.
\newblock
  \href{https://dx.doi.org/https://doi.org/10.1103/PhysRevLett.122.180501}{Physical
  review letters}~(2019).
\newblock  \href{http://arxiv.org/abs/1807.02467}{arXiv:1807.02467}.

\bibitem{tom:error}
Thomas~E O'Brien, Stefano Polla, Nicholas~C Rubin, William~J Huggins, Sam
  McArdle, Sergio Boixo, Jarrod~R McClean, and Ryan Babbush.
\newblock ``Error mitigation via verified phase estimation''.
\newblock
  \href{https://dx.doi.org/https://doi.org/10.1103/PRXQuantum.2.020317}{Physical
  review X Quantum}~(2021).
\newblock  \href{http://arxiv.org/abs/2010.02538}{arXiv:2010.02538}.

\bibitem{ubaru:qtda}
Shashanka Ubaru, Ismail~Yunus Akhalwaya, Mark~S Squillante, Kenneth~L Clarkson,
  and Lior Horesh.
\newblock ``Quantum topological data analysis with linear depth and exponential
  speedup''~(2021).
\newblock  \href{http://arxiv.org/abs/2108.02811}{arXiv:2108.02811}.

\bibitem{berry:hs}
Dominic~W Berry, Andrew~M Childs, and Robin Kothari.
\newblock ``Hamiltonian simulation with nearly optimal dependence on all
  parameters''.
\newblock
  \href{https://dx.doi.org/https://doi.org/10.1145/3313276.3316386}{Proceedings
  of 56th Annual Symposium on Foundations of Computer Science}~(2015).
\newblock  \href{http://arxiv.org/abs/1501.01715}{arXiv:1501.01715}.

\bibitem{rennela:counter}
Mathys Rennela, Alfons Laarman, and Vedran Dunjko.
\newblock ``Hybrid divide-and-conquer approach for tree search
  algorithms''~(2020).
\newblock  \href{http://arxiv.org/abs/2007.07040}{arXiv:2007.07040}.

\bibitem{dutkiewicz:qpe}
Alicja Dutkiewicz, Barbara~M Terhal, and Thomas~E O'Brien.
\newblock ``Heisenberg-limited quantum phase estimation of multiple eigenvalues
  with a single control qubit''~(2021).
\newblock  \href{http://arxiv.org/abs/2107.04605}{arXiv:2107.04605}.

\bibitem{tom:qpe}
Thomas~E. O'Brien, Brian Tarasinski, and Barbara Terhal.
\newblock ``Quantum phase estimation of multiple eigenvalues for small-scale
  (noisy) experiments''.
\newblock
  \href{https://dx.doi.org/https://doi.org/10.1088/1367-2630/aafb8e}{New
  Journal of Physics}~(2019).
\newblock  \href{http://arxiv.org/abs/1809.09697}{arXiv:1809.09697}.

\bibitem{somma:qpe}
Rolando~D Somma.
\newblock ``Quantum eigenvalue estimation via time series analysis''.
\newblock
  \href{https://dx.doi.org/https://doi.org/10.1088/1367-2630/ab5c60}{New
  Journal of Physics}~(2019).
\newblock  \href{http://arxiv.org/abs/1907.11748}{arXiv:1907.11748}.

\bibitem{kato:perturbation}
Tosio Kato.
\newblock ``Perturbation theory for linear operators''.
\newblock
  \href{https://dx.doi.org/https://doi.org/10.1007/978-3-642-66282-9}{Volume
  132}.
\newblock Springer Science \& Business Media. ~(2013).

\bibitem{feng:rank}
Xinlong Feng and Zhinan Zhang.
\newblock ``The rank of a random matrix''.
\newblock
  \href{https://dx.doi.org/https://doi.org/10.1016/j.amc.2006.07.076}{Applied
  mathematics and computation {\bf 185}, 689--694}~(2007).

\bibitem{huang:demonstration}
He-Liang Huang, Xi-Lin Wang, Peter~P Rohde, Yi-Han Luo, You-Wei Zhao, Chang
  Liu, Li~Li, Nai-Le Liu, Chao-Yang Lu, and Jian-Wei Pan.
\newblock ``Demonstration of topological data analysis on a quantum
  processor''.
\newblock
  \href{https://dx.doi.org/https://doi.org/10.1364/OPTICA.5.000193}{Optica {\bf
  5}, 193--198}~(2018).
\newblock  \href{http://arxiv.org/abs/1801.06316}{arXiv:1801.06316}.

\end{thebibliography}

\clearpage



\section*{Supplementary material}

# Supplemental Material – Towards Quantum Advantage via Topological Data Analysis

## I. LLSD IS DQC1-HARD

Following the definition of [1], for any problem $L \in \mathsf{DQC1}$ and every $x \in L$, there exists a quantum circuit $U$ of depth $T \in \mathcal{O}\left(\operatorname{poly}(|x|)\right)$ that operates on $n \in \mathcal{O}\left(\operatorname{poly}(|x|)\right)$ qubits such that

- $x \in L_{yes} \implies p_0 \geq \frac{1}{2} + \frac{1}{\operatorname{poly}(|x|)}$,
- $x \in L_{no} \implies p_0 \leq \frac{1}{2} - \frac{1}{\operatorname{poly}(|x|)}$,

where $p_0 = \operatorname{Tr}\left[(|0\rangle\langle 0| \otimes I) U \rho U^\dagger\right]$ and $\rho = |0\rangle\langle 0| \otimes I/2^{n-1}$. From this it can be gathered that if we can estimate $p_0$ to within $1/\operatorname{poly}(|x|)$ additive precision, then we can solve $L$.

For a positive semidefinite matrix $H \in \mathbb{C}^{2^n \times 2^n}$ and a threshold $b \in \mathbb{R}_{\geq 0}$, we define the *normalized subtrace* of $H$ up to $b$ as

$$\overline{\operatorname{Tr}_b}(H) = \frac{1}{2^n} \sum_{0 \leq \lambda_k \leq b} \lambda_k,$$

where $\lambda_0 \leq \cdots \leq \lambda_{2^n-1}$ denote the eigenvalues of $H$. The following result by Brandão shows that if we can estimate the normalized subtrace $\overline{\operatorname{Tr}_b}$ of log-local Hamiltonians up to additive inverse polynomial precision, then we can solve any problem in $\mathsf{DQC1}$. In other words, estimating $\overline{\operatorname{Tr}_b}$ of log-local Hamiltonians up to additive inverse polynomial precision is $\mathsf{DQC1}$-hard.

**Proposition 1** (Brandão [2]). *Given as input a description of an $n$-qubit quantum circuit $U$ of depth $T \in \mathcal{O}\left(\operatorname{poly}(n)\right)$ together with a polynomial $r(n)$, one can efficiently construct a log-local Hamiltonian $H \in \mathbb{C}^{T2^n \times T2^n}$ and a threshold $b \in \mathcal{O}\left(\operatorname{poly}(n)\right)$ such that*

$$\left|\overline{\operatorname{Tr}_b}(H) - p_0\right| \leq \frac{1}{r(n)}, \tag{1}$$

*where $p_0 = \operatorname{Tr}\left[(|0\rangle\langle 0| \otimes I) U \rho U^\dagger\right]$ and $\rho = |0\rangle\langle 0| \otimes I/2^{n-1}$. Moreover, $H$ also satisfies:*

(i) *$H$ is positive semidefinite.*

(ii) *There exists a $\delta \in \Omega\left(1/\operatorname{poly}(n)\right)$ such that $H$ has no eigenvalues in the interval $[b, b+\delta]$.*

**Remark.** *The Hamiltonian in the above proposition is obtained by applying Kitaev's circuit-to-Hamiltonian construction directly to the circuit $U$, but only constraining the input and output of the clean qubit while leaving the other qubits unconstrained (emulating the maximally mixed state).*

We will show that we can efficiently estimate the normalized subtrace $\overline{\operatorname{Tr}_b}$ in Equation 1 to within additive inverse polynomial precision using an oracle for LLSD. To be precise, we show that we can estimate this normalized subtrace to within additive inverse polynomial precision using a polynomial amount of nonadaptive queries to an oracle for LLSD (whose input is restricted to log-local Hamiltonians), together with polynomial-time classical preprocessing of the inputs and postprocessing of the outputs. In other words, we provide a polynomial-time truth-table reduction from the problem of estimating $\overline{\operatorname{Tr}_b}$ to LLSD. We gather this in Lemma 2, which together with Proposition 1 shows that LLSD with the input restricted to log-local Hamiltonians is $\mathsf{DQC1}$-hard under polynomial-time truth-table reductions.

**Lemma 2.** *Given as input $H \in \mathbb{C}^{T2^n \times T2^n}$ and $b \in \mathcal{O}\left(\operatorname{poly}(n)\right)$ as described in Proposition 1, together with a polynomial $q(n)$, one can compute a quantity $\Lambda$ that satisfies*

$$|\Lambda - \overline{\operatorname{Tr}_b}(H)| \leq \frac{1}{q(n)},$$

*using a polynomial number of queries to an oracle for LLSD, together with polynomial-time classical preprocessing of the inputs and postprocessing of the outputs.*

*Proof.* Define $\Delta = (3q(n))^{-1}$, $M = b/\Delta$, $\epsilon = (6Mbq(n))^{-1}$ and let $\delta < \Delta/3$ be such that $H$ has no eigenvalues in the interval $[b, b+\delta]$. Also, define the thresholds $x_j = (j+1)\Delta$, for $j = 0, \ldots, M-1$. Next, denote by $\hat{\chi}_j$ the outcome of LLSD with threshold $b = x_j$ and precision parameters $\delta, \epsilon$ as defined above. That is, $\hat{\chi}_j$ is an estimate of $\hat{y}_j$ to within additive accuracy $\epsilon$, where

$$\hat{y}_j = N_H(0, x_j) + \hat{\gamma}_j, \text{ with } 0 \leq \hat{\gamma}_j \leq N_H(x_j, x_j + \delta).$$

Subsequently, define $\chi_0 = \hat{\chi}_0$, $y_0 = \hat{y}_0$ and

$$y_j = \hat{y}_j - \hat{y}_{j-1}, \tag{2}$$
$$\chi_j = \hat{\chi}_j - \hat{\chi}_{j-1}, \tag{3}$$

for $1 \leq j \leq M-1$. Finally, define the estimate

$$\Lambda = \sum_{j=0}^{M-1} \chi_j x_j. \tag{4}$$

We will show that $\Lambda$ is indeed an estimate of $\overline{\text{Tr}_b}(H)$ to within additive precision $\pm 1/q(n)$. To do so, we define $\gamma_0 = \hat{\gamma}_0$ and $\gamma_j = \hat{\gamma}_j - \hat{\gamma}_{j-1}$ for $1 \leq j \leq M-1$, and we define and expand

$$\Gamma = \sum_{j=0}^{M-1} y_j x_j = \sum_{j=0}^{M-1} \left(N_H(x_{j-1}, x_j) + \gamma_j\right) x_j = \underbrace{\sum_{j=0}^{M-1} N_H(x_{j-1}, x_j) x_j}_{\mathcal{B}:=} + \underbrace{\sum_{j=0}^{M-1} \gamma_j x_j}_{\mathcal{E}_{bin}:=}.$$

We start by upper-bounding the magnitude of the $\mathcal{E}_{bin}$ term. To do so, we rewrite

$$\mathcal{E}_{bin} = \sum_{j=0}^{M-1} \gamma_j x_j = \hat{\gamma}_0 x_0 + \sum_{j=1}^{M-1} (\hat{\gamma}_j - \hat{\gamma}_{j-1}) x_j$$
$$= \sum_{j=0}^{M-1} \hat{\gamma}_j x_j - \sum_{j=1}^{M-1} \hat{\gamma}_{j-1} x_j$$
$$= \sum_{j=0}^{M-1} \hat{\gamma}_j x_j - \sum_{j=1}^{M-1} \hat{\gamma}_{j-1}(x_{j-1} + \Delta)$$
$$= \sum_{j=0}^{M-1} \hat{\gamma}_j x_j - \sum_{j=1}^{M-1} \hat{\gamma}_{j-1} x_{j-1} - \Delta \sum_{j=1}^{M-1} x_{j-1}$$
$$= \underbrace{\hat{\gamma}_{M-1}}_{=0} x_{M-1} - \Delta \underbrace{\sum_{j=1}^{M-1} \hat{\gamma}_{j-1}}_{\leq 1},$$

and we conclude that $|\mathcal{E}_{bin}| \leq \Delta$. Next, we upper-bound the absolute difference of $\mathcal{B}$ and $\overline{\text{Tr}_b}(H)$.

$$\left|\mathcal{B} - \overline{\text{Tr}_b}(H)\right| = \left|\sum_{j=0}^{M-1} N_H(x_{j-1}, x_j) x_j - \overline{\text{Tr}_b}(H)\right| \leq \sum_{j=0}^{M-1} \Delta \cdot N_H(x_{j-1}, x_j) \leq \Delta.$$

Finally, we upper-bound the absolute difference between $\Lambda$ and $\Gamma$.

$$|\Lambda - \Gamma| = \left|\sum_{j=0}^{M-1} (\chi_j - y_j) x_j\right| \leq \left|\sum_{j=0}^{M-1} 2\epsilon x_j\right| \leq M \cdot 2\epsilon \cdot b = \frac{1}{3q(n)}$$

Combining all of the above we find that

$$|\Lambda - \overline{\text{Tr}_b}(H)| \leq |\Lambda - \Gamma| + |\Gamma - \overline{\text{Tr}_b}(H)| \leq |\Lambda - \Gamma| + |\mathcal{B} - \overline{\text{Tr}_b}(H)| + |\mathcal{E}| \leq \frac{1}{3q(n)} + \Delta + \Delta = \frac{1}{q(n)}.$$

□

## II. QUANTUM ALGORITHMS FOR SUES AND LLSD

In this section we give a quantum algorithm for SUES and a quantum algorithm for LLSD. Moreover, if the input is a log-local Hamiltonian, then the quantum algorithms we give in this section turn out to be a DQC1 algorithm in the case of LLSD, and a $\mathsf{DQC1}_{\log n}$ algorithm in the case of SUES. That is, if the input is a log-local Hamiltonian, then these algorithms can be implemented in the one clean qubit model, where in the case of SUES we need to measure logarithmically many qubits (as opposed to just one), in order to read out the entire encoding of the eigenvalue.

By scaling the input $H' = H/\Lambda$, where $\Lambda \in \mathcal{O}(\text{poly}(n))$ is an upper bound on the largest eigenvalue of $H$, we can assume without loss of generality that $||H|| < 1$. Moreover, we will use that allowing up to $\mathcal{O}(\log(n))$ clean qubits does not change the class DQC1 [1]. That is, the class of problems that can be solved in polynomial time using the one clean qubit model of computation is the same as the class of problems that can be solved in polynomial time using the $k$-clean qubit model of computation, for $k \in \mathcal{O}(\log n)$. We use this result since the quantum algorithms we describe need additional ancilla qubits, which have to be initialized in the all-zeros state and hence be 'clean'.

### A. Quantum algorithm for SUES

In this section we describe a quantum algorithm for SUES, which when the input is restricted to log-local Hamiltonians turns out to be a $\mathsf{DQC1}_{\log n}$ algorithm. That is, if the input is a log-local Hamiltonian, then this algorithm can be implemented using the one clean qubit model of computation where we are allowed to measure logarithmically many of the qubits at the end, in order to read out the encoding of the eigenvalue.

The quantum algorithm for SUES implements an approximation of the unitary $e^{iH}$ using Hamiltonian simulation, to which it applies quantum phase estimation with the eigenvector register starting out in the maximally mixed state. In the remainder of this section we will show that we can control the errors such that quantum phase estimation applied to the approximation of $e^{iH}$ outputs the corresponding eigenvalue of $H$ up to precision $\delta \in \Omega(1/\text{poly}(n))$, with error probability $\mu \in \Omega(1/\text{poly}(n))$. Because the maximally mixed state is in a given eigenstate with uniform probabilities over all eigenstates, this shows that this quantum algorithm is able to output a sample from a $(\delta, \mu)$-approximation of the uniform distribution over the eigenvalues of $H$.

Errors can arise in two places, namely due to the imprecisions of the unitary implemented by the Hamiltonian simulation and due to the imprecisions of estimating eigenvalues using quantum phase estimation. First, we discuss the errors of the Hamiltonian simulation step. Given sparse access to $H$, we can implement a unitary $V$ such that

$$||V - e^{iH}|| < \gamma, \tag{5}$$

in time $\mathcal{O}(\text{poly}(n, \log(1/\gamma)))$ [3]. The algorithms for Hamiltonian simulation of matrices specified by an oracle unfortunately require more than $\mathcal{O}(\log n)$ ancilla qubits, which implies that they can not be implemented using the one clean qubit model. On the other hand, if $H$ is a log-local Hamiltonian, then Hamiltonian simulation techniques based on the Trotter-Suzuki formula can implement a unitary $V$ that satisfies Equation 5 in time $\mathcal{O}(\text{poly}(n, 1/\gamma))$ [4], while only using a constant number of ancilla qubits [5]. Therefore, if $H$ is a log-local Hamiltonian, then using the one clean qubit model we can implement a unitary $V$ that satisfies Equation 5 in time $\mathcal{O}(\text{poly}(n, 1/\gamma))$.

Denote by $\lambda_j$ and $\zeta_j$ the output of the quantum phase estimation routine (where for now we assume that it works perfectly, i.e., introduces no error) when run using $e^{iH}$ and $V$, respectively. Then, by Equation 5 we have

$$|e^{i\lambda_j} - e^{i\zeta_j}| \leq \gamma,$$

where we assume that $|\lambda_j - \zeta_j| \leq \pi$ by adding multiples of $2\pi$ to $\lambda_j$ if necessary. With some algebra [5], we can show that this implies that

$$|\lambda_j - \zeta_j| \leq \pi\gamma/2.$$

Choosing the accuracy of the Hamiltonian simulation to be $\gamma = \delta/\pi \in \Omega(1/\text{poly}(n))$, we get that

$$|\lambda_j - \zeta_j| < \delta/2. \tag{6}$$

Next, we will consider the errors that arise from using the quantum phase estimation routine to estimate the eigenvalues $\zeta_j$ of the unitary $V$. The quantum phase estimation routine requires a register of $t$ ancilla qubits (also called the eigenvalue register), onto which the eigenvalue will be loaded. If we take

$$t = \log(2/\delta) + \lceil \log(2 + 1/2\mu) \rceil \in \mathcal{O}(\log n)$$

qubits in the eigenvalue register, then quantum phase estimation outputs an estimate $\overline{\zeta_j}$ that satisfies

$$|\overline{\zeta_j} - \zeta_j| \leq \delta/2,$$

with probability at least $(1-\mu)$ [6]. In particular, with probability at least $(1-\mu)$ this estimate satisfies

$$|\overline{\zeta_j} - \lambda_j| \leq |\overline{\zeta_j} - \zeta| + |\zeta_j - \lambda_j| \leq \delta.$$

This requires $\widetilde{\mathcal{O}}(2^t) = \widetilde{\mathcal{O}}(\text{poly}(n))$ applications of the unitary $V$, each of which can be implemented in $\mathcal{O}(\text{poly}(n))$ time as discussed above. In addition, this quantum phase estimation step requires only $\mathcal{O}(\log n)$ ancilla qubits, making it possible to be implemented using the one clean qubit model.

In conclusion, both the Hamiltonian simulation and the quantum phase estimation can be implemented up to the required precision in time $\mathcal{O}(\text{poly}(n))$. Moreover, if $H$ is a log-local Hamiltonian, then this can be done using the one clean qubit model. Finally, to read out the encoding of the eigenvalue, we need to measure the $t \in \mathcal{O}(\log(n))$ qubits in the eigenvalue register, resulting in a $\mathsf{DQC1}_{\log n}$ algorithm for SUES if the input is a log-local Hamiltonian.

### B. Quantum algorithm for LLSD

In this section, we will describe two quantum algorithms for LLSD, both of which turn into $\mathsf{DQC1}$ algorithms when the input is restricted to log-local Hamiltonians. That is, if the input is a log-local Hamiltonian, then these algorithms can be implemented using the one clean qubit model of computation.

#### 1. Counting eigenvalues below the threshold

A straightforward approach is to solving LLSD is to repeatedly sample from the output of SUES and then compute the fraction of samples that lie below the given threshold. The downside of this is that it requires one to measure the entire eigenvalue register consisting of logarithmically many qubits, which is prohibitive as we are only allowed to measure a single qubit in the one clean qubit model. This can be circumvented by simply adding an extra clean qubit and flipping this qubit conditioned on the state in the eigenvalue register being smaller than the given threshold. This extra qubit will be flipped with probability close to the low-lying spectral density, allowing us to obtain a solution to LLSD by only measuring this single qubit. Moreover, if $H$ is a log-local Hamiltonian, then this 'fully quantum' algorithm can be implemented using the one clean qubit model, as it requires only a few more additional clean qubits on top of those required for the quantum algorithm for SUES discussed in Section II A.

Note that the outcome probabilities of this 'fully quantum' algorithm are identical to those obtained by measuring the entire eigenvalue register, followed by classical counting of the number of samples below the given threshold. Consequently, the same error analysis applies in both cases. In the rest of this section we will discuss the error analysis of classically counting the number of samples below the given threshold.

Let $m = \epsilon^{-2} \in \mathcal{O}(\text{poly}(n))$ and draw for $j = 1, \ldots, m$ a sample $\overline{\lambda}_{k_j}$ from SUES with $\delta/2$ as the precision parameter. Next, compute

$$\chi_j = \begin{cases} 1 \text{ if } \overline{\lambda}_{k_j} \in (a - \delta/2, b + \delta/2), \\ 0 \text{ otherwise.} \end{cases}$$

For now we assume that all samples $\overline{\lambda}_{k_j}$ were *correctly sampled*, i.e., each $k_j$ is drawn uniformly at random from the set $\{0, \ldots, 2^n - 1\}$ and $|\lambda_{k_j} - \overline{\lambda}_{k_j}| \leq \delta/2$, where $\lambda_{k_j}$ denotes the eigenvalue of which $\overline{\lambda}_{k_j}$ is an estimate. We now show that under this assumption the quantity

$$\chi := \frac{1}{m} \sum_{j=1}^{m} \chi_j \tag{7}$$

is, with high probability, a correct solution to LLSD. By the Chernoff-Hoeffding inequality $\chi$ is, with high probability, an estimate to within additive precision $\epsilon$ of

$$y := \Pr_{\overline{\lambda} \sim \text{SUES}}\left[\overline{\lambda} \in (a - \delta/2, b + \delta/2)\right],$$

where the probability is taken over the $\overline{\lambda}$ being correctly sampled from SUES. Because we assume that the $\overline{\lambda}$ are correctly samples from SUES, we know that they satisfy $|\lambda - \overline{\lambda}| \leq \delta/2$, where $\lambda$ denotes the eigenvalue of which $\overline{\lambda}$ is an estimate. This implies that

(i) $y \leq \Pr_{\lambda \sim_U \{\lambda_j\}_{j=1}^{2^n}} \left[ \lambda \in (a - \delta, b + \delta) \right] = N_H(a - \delta, b + \delta)$,

(ii) $y \geq \Pr_{\lambda \sim_U \{\lambda_j\}_{j=1}^{2^n}} \left[ \lambda \in (a, b) \right] = N_H(a, b)$,

where the probabilities are taken over the $\lambda$ being sampled uniformly from the set of all eigenvalues of $H$. Combining this with the Chernoff-Hoeffding inequality, we find that $\chi$ is, with high probability, an estimate of $y$ up to additive precision $\epsilon$, where $y$ satisfies

$$N_H(a, b) \leq y \leq N_H(a - \delta, b + \delta).$$

That is, if all $\overline{\lambda}_{k_j}$ were sampled correctly from SUES, then $\chi$ is with high probability a correct solution to LLSD.

Finally, we consider the probability that all samples $\overline{\lambda}_{k_j}$ were indeed sampled correctly. By the union bound this probability is at least $1 - m\mu$, where $\mu$ denotes the sampling error probability of SUES. Because $m \in \mathcal{O}(\mathrm{poly}(n))$, we can choose $\mu \in \Omega\left(1/\mathrm{poly}(\epsilon^{-2}, n)\right) = \Omega\left(1/\mathrm{poly}(n)\right)$ such that all our samples are sampled correctly with probability close to 1. Therefore, we conclude that the $\chi$ defined in Equation 7 is a correct solution to LLSD, with probability close to 1. Moreover, $\chi$ can be obtained from a polynomial number of samples from SUES, and can therefore be computed in time $\mathcal{O}(\mathrm{poly}(n))$.

### 2. Using trace estimation of eigenvalue transform

In our paper, we use a result of Cade & Montanaro [5] to argue that the complexity of estimating the spectral entropy of a Hermitian matrix is closely related to DQC1. In their work, Cade & Montanaro describe a DQC1 algorithm can estimate traces of general functions of Hermitian matrices (i.e., beyond spectral entropies). This algorithm could also be used to extract other interesting properties encoded in the spectrum of the combinatorial Laplacian. To illustrate this and connect even further to this line of work, we provide an alternative algorithm for LLSD based on this algorithm. The main result we will utilize is the following Lemma.

**Lemma 3** (Cade & Montanaro [5]). *For a log-local Hamiltonian $H \in \mathbb{C}^{2^n \times 2^n}$, and any log-space polynomial-time computable function $f: I \to [-1, 1]$ (where $I$ contains the spectrum of $H$) that is Lipschitz continuous with constant $K$ (i.e., $|f(x) - f(y)| \leq K|x - y|$ for all $x, y \in I$), there exists a DQC1 algorithm to estimate $\mathrm{Tr}(f(H))/2^n = \sum_j f(\lambda_j)/2^n$ up to additive accuracy $\epsilon(K+1)$, where $\lambda_j$ denote the eigenvalues of $H$, and $\epsilon \in \Omega(1/poly(n))$.*

It is clear that if the function $f$ is the step-function with threshold $b + \delta/2$ given by

$$f(x) = \begin{cases} 1 & \text{if } x \leq b + \delta/2, \\ 0 & \text{otherwise,} \end{cases}$$

then the quantity estimated by the algorithm of Lemma 3 is a correct solution to LLSD. However, as this function is not Lipschitz continuous, we will use a smooth approximation based on the following lemma.

**Lemma 4** (Smooth approximation of the sign function). *Let $\delta > 0$, $\epsilon \in (0, 1)$ and $\gamma = \frac{\delta \sqrt{2\epsilon - \epsilon^2}}{1 - \epsilon}$. Then, the function $g_\gamma(x) = \frac{x}{\sqrt{x^2 + \gamma^2}}$ satisfies*

(i) *for all $x \in [-2, 2]$ : $-1 \leq g_\gamma(x) \leq 1$,*

(ii) *for all $x \in [-2, 2] \setminus (-\delta, \delta)$ : $|g_\gamma(x) - sgn(x)| \leq \epsilon$, and*

(iii) $\sup_{x \in [-2, 2]} |g'_\gamma(x)| \leq \frac{1}{\gamma}$.

*Proof.* (i) It is clear that for all $x \in [-2, 2]$ we have: $-1 \leq g_\gamma(-2) \leq g_\gamma(x) \leq g_\gamma(2) \leq 1$.

(ii) Let $x \in (\delta, 2]$, then

$$|g_\gamma(x) - \mathrm{sgn}(x)| = |g_\gamma(x) - 1| \leq |g_\gamma(\delta) - 1| = \epsilon.$$

For $x \in [-2, -\delta)$ we note that

$$|g_\gamma(x) - \mathrm{sgn}(x)| = |g_\gamma(x) + 1| \leq |g_\gamma(-\delta) + 1| = |-(g_\gamma(\delta) - 1)| = \epsilon.$$

(*iii*) It is clear that: $\sup_{x \in [-2,2]} |g'_\gamma(x)| = |g'_\gamma(0)| = \frac{1}{\gamma}$.

□

Let $\gamma = \frac{(\delta/2)\sqrt{2\epsilon - \epsilon^2}}{1-\epsilon}$ and define $g = g_\gamma$ as in Lemma 4. We define our smooth approximation of the step-function by

$$\hat{f}(x) = \frac{g(-x + b') + 1}{2},$$

where $b' = b + \delta/2$. By Lemma 4 we know that $\hat{f}$ is Lipschitz continuous on $[0, 1]$ with constant $1/\gamma \in \mathcal{O}(\text{poly}(n))$, and that it satisfies

- for all $x \in [0, 1] : 0 \leq \hat{f}(x) \leq 1$, and
- for all $x \in [0, 1] \setminus (b, b + \delta) : |\hat{f}(x) - f(x)| \leq \epsilon/2$.

Subsequently, we define our estimation objective

$$y = \frac{1}{2^n} \left( \sum_{j \,:\, \lambda_j \in [0,b]} f(\lambda_j) + \sum_{j \,:\, \lambda_j \in [b, b+\delta]} \hat{f}(\lambda_j) \right),$$

and we note that $y$ indeed satisfies $N_H(0, b) \leq y \leq N_H(0, b + \delta)$, since

$$y = \frac{1}{2^n} \left( \sum_{j \,:\, \lambda_j \in [0,b]} f(\lambda_j) + \sum_{j \,:\, \lambda_j \in [b, b+\delta]} \hat{f}(\lambda_j) \right)$$

$$= N_H(0, b) + \underbrace{\frac{1}{2^n} \sum_{j \,:\, \lambda_j \in [b, b+\delta]} \underbrace{\hat{f}(\lambda_j)}_{\in [0,1]}}_{\in [0, N_H(b, b+\delta)]}.$$

Now our goal is to use Lemma 3 to obtain an $\epsilon$-approximation of $y$. To this end, we first define

$$\Lambda = \frac{1}{2^n} \sum_{j=1}^{2^n} \hat{f}(\lambda_j),$$

and we upper-bound the absolute difference between $y$ and $\Lambda$ as follows

$$\left| \Lambda - y \right| \leq \frac{1}{2^n} \left| \sum_{j \,:\, \lambda_j \in [0,b]} \left( \hat{f}(\lambda_j) - f(\lambda_j) \right) + \sum_{j \,:\, \lambda_j \in [b+\delta, 1]} \hat{f}(\lambda_j) \right|$$

$$\leq \frac{1}{2^n} \left( \sum_{j \,:\, \lambda_j \in [0,b]} \left| \hat{f}(\lambda_j) - f(\lambda_j) \right| + \sum_{j \,:\, \lambda_j \in [b+\delta, 1]} \left| \hat{f}(\lambda_j) \right| \right)$$

$$\leq \frac{1}{2^n} \left( \sum_{j \,:\, \lambda_j \in [0,b]} \epsilon/2 + \sum_{j \,:\, \lambda_j \in [b+\delta, 1]} \epsilon/2 \right) \leq \epsilon/2.$$

Finally, let $\chi$ be the output of the algorithm of Lemma 3 applied to our function $\hat{f}$ with precision parameter $\hat{\epsilon} = \epsilon/(2(K+1)) \in \Omega(1/\text{poly}(n))$. In particular, $\chi$ satisfies $|\chi - \Lambda| \leq \hat{\epsilon}(K+1) = \epsilon/2$. We conclude that $\chi$ is a correct solution to LLSD since

$$|\chi - y| \leq |\chi - \Lambda| + |\Lambda - y| \leq \epsilon/2 + \epsilon/2 = \epsilon,$$

and $y$ indeed satisfies $N_H(0, b) \leq y \leq N_H(0, b + \delta)$ as discussed earlier.

## III. SWES IS DQC1-HARD

In this section, we will show that SWES is DQC1-hard. We will do so by showing that we estimate the DQC1-hard normalized subtrace $\overline{\mathrm{Tr}}_b(H)$ from Proposition 1 up to additive polynomial precision $\epsilon \in \Omega(1/\mathrm{poly})$ using a polynomial number of queries to an oracle for SWES, together with polynomial-time classical preprocessing of the input and postprocessing of the output.

First, by considering how $H$ is constructed in [2], we note that $\mathrm{Tr}(H)$ is known and that $\mathrm{Tr}(H)/2^n \in \mathcal{O}(\mathrm{poly}(n))$. Next, we define $\hat{\epsilon} = (\epsilon/(\mathrm{Tr}(H)/2^n))$ and $m = 1/\hat{\epsilon}^2$. Subsequently, let $\overline{\lambda}_{k_1}, \ldots, \overline{\lambda}_{k_m}$ denote samples drawn from SWES with estimation precision $\delta/2$, where $\delta$ is such that $H$ has no eigenvalues in $[b, b+\delta]$. For now we assume that all samples were *correctly sampled*, i.e., $|\overline{\lambda}_{k_j} - \lambda_{k_j}| \leq \delta/2$, where $\lambda_{k_j}$ denotes the eigenvalue of which $\overline{\lambda}_{k_j}$ is an estimate. Afterwards, we estimate the normalized subtrace $\overline{\mathrm{Tr}}_b(H)$ by computing the ratio of samples that is below $b + \delta/2$

$$\chi = \frac{1}{m} \sum_{j \,:\, \overline{\lambda}_{k_j} \leq b+\delta/2} 1$$

By the Chernoff-Hoeffding inequality (together with the fact that $H$ has no eigenvalues in $[b, b+\delta]$), this ratio $\chi$ is, with high probability, an estimate of

$$\Lambda = \sum_{j \,:\, \lambda_j \leq b} \lambda_j / \mathrm{Tr}(H),$$

up to additive precision $\hat{\epsilon}$. Therefore, $(\mathrm{Tr}(H)/2^n) \cdot \chi$ is, with high probability, an $\epsilon$ estimate of $(\mathrm{Tr}(H)/2^n) \cdot \Lambda = \overline{\mathrm{Tr}}_b(H)$.

Finally, we consider the probability that all samples $\overline{\lambda}_{k_j}$ were indeed sampled correctly. By the union bound this probability is $M \cdot \mu$, where $\mu$ denotes the sampling error probability of SWES. Because $m \in \mathcal{O}(\mathrm{poly}(n))$, we can choose $\mu \in \Omega(1/m) = \mathcal{O}(\mathrm{poly}(n))$ such that all our samples are sampled correctly with probability close to 1.

---

[1] P. W. Shor and S. P. Jordan, Quantum Information & Computation **8**, 681 (2008), `arXiv:0707.2831`.
[2] F. G. Brandão, *Entanglement theory and the quantum simulation of many-body physics*, Ph.D. thesis, University of London (2008), `arXiv:0810.0026`.
[3] G. H. Low and I. L. Chuang, Physical review letters **118**, 010501 (2017), `arXiv:1606.02685`.
[4] S. Lloyd, Science , 1073 (1996).
[5] C. Cade and A. Montanaro, in *13th Conference on the Theory of Quantum Computation, Communication and Cryptography* (2018) `arXiv:1706.09279`.
[6] M. A. Nielsen and I. L. Chuang, *Quantum Computation and Quantum Information: 10th Anniversary Edition*, 10th ed. (Cambridge University Press, 2011).



\end{document}